\documentclass[acmsmall]{acmart}
\AtBeginDocument{%
  \providecommand\BibTeX{{%
    \normalfont B\kern-0.5em{\scshape i\kern-0.25em b}\kern-0.8em\TeX}}}

\setcopyright{acmcopyright}
\copyrightyear{2018}
\acmYear{2018}
\acmDOI{XXXXXXX.XXXXXXX}

\acmJournal{TIST}
\acmVolume{37}
\acmNumber{4}
\acmArticle{111}
\acmMonth{8}

\usepackage[at]{easylist}
\usepackage{caption}
\usepackage{subcaption}

\usepackage{nomencl}
\makenomenclature

\nomenclature{AMC}{automatic modulation classification}
\nomenclature{NN}{neural network}
\nomenclature{FO}{frequency offset}
\nomenclature{CNN}{convolutional neural network}
\nomenclature{RRC}{root-raised cosine}
\nomenclature{CV}{computer vision}
\nomenclature{ML}{machine learning}
\nomenclature{TL}{transfer learning}
\nomenclature{RF}{radio frequency}
\nomenclature{DL}{deep learning}
\nomenclature{NLP}{natural language processing}
\nomenclature{IQ}{in-phase/quadrature}
\nomenclature{SNR}{signal-to-noise ratio}
\nomenclature{AWGN}{additive white Gaussian noise}
\nomenclature{LEEP}{Log Expected Empirical Prediction}
\nomenclature{LogME}{Logarithm of Maximum Evidence}

\nomenclature{BPSK}{binary phase-shift keying}
\nomenclature{QPSK}{quadrature phase-shift keying}
\nomenclature{PSK8}{phase-shift keying, order 8}
\nomenclature{PSK16}{phase-shift keying, order 16}
\nomenclature{OQPSK}{offset quadrature phase-shift keying}
\nomenclature{QAM16}{quadrature amplitude modulation, order 16}
\nomenclature{QAM32}{quadrature amplitude modulation, order 32}
\nomenclature{QAM64}{quadrature amplitude modulation, order 64}
\nomenclature{APSK16}{amplitude and phase-shift keying, order 16}
\nomenclature{APSK32}{amplitude and phase-shift keying, order 32}
\nomenclature{FSK5k}{frequency-shift keying, 5kHz carrier spacing}
\nomenclature{FSK75k}{frequency-shift keying, 75kHz carrier spacing}
\nomenclature{GFSK5k}{Gaussian frequency-shift keying, 5kHz carrier spacing}
\nomenclature{GFSK75k}{Gaussian frequency-shift keying, 75kHz carrier spacing}
\nomenclature{MSK}{minimum-shift keying}
\nomenclature{GMSK}{Gaussian minimum-shift keying}
\nomenclature{FM-NB}{narrow band frequency modulation}
\nomenclature{FM-WB}{wide band frequency modulation}
\nomenclature{AM-DSB}{amplitude modulation, double-sideband}
\nomenclature{AM-DSBSC}{amplitude modulation, double-sideband suppressed-carrier}
\nomenclature{AM-LSB}{amplitude modulation, lower-sideband}
\nomenclature{AM-USB}{amplitude modulation, upper-sideband}

\usepackage{acro}
\DeclareAcronym{AMC}{
  short = AMC ,
  long  = automatic modulation classification ,
  tag = abbrev
}

\DeclareAcronym{NN}{
  short = NN ,
  long  = neural network ,
  tag = abbrev
}

\DeclareAcronym{FO}{
  short = FO ,
  long  = frequency offset ,
  tag = abbrev
}

\DeclareAcronym{MLP}{
  short = MLP ,
  long  = multi-layer perceptron ,
  tag = abbrev
}

\DeclareAcronym{CNN}{
  short = CNN ,
  long  = convolutional neural network ,
  tag = abbrev
}

\DeclareAcronym{RNN}{
  short = RNN ,
  long  = recurrent neural network ,
  tag = abbrev
}

\DeclareAcronym{RRC}{
  short = RRC ,
  long  = root-raised cosine ,
  tag = abbrev
}

\DeclareAcronym{LSTM}{
  short = LSTM ,
  long  = long short-term memory ,
  tag = abbrev
}

\DeclareAcronym{UAV}{
  short = UAV ,
  long  = unmanned aerial vehicle ,
  tag = abbrev
}

\DeclareAcronym{CV}{
  short = CV ,
  long  = computer vision ,
  tag = abbrev
}

\DeclareAcronym{ML}{
  short = ML ,
  long  = machine learning ,
  tag = abbrev
}

\DeclareAcronym{TL}{
  short = TL ,
  long  = transfer learning ,
  tag = abbrev
}

\DeclareAcronym{AE}{
  short = AE ,
  long  = autoencoder ,
  tag = abbrev
}

\DeclareAcronym{RF}{
  short = RF ,
  long  = radio frequency ,
  tag = abbrev
}

\DeclareAcronym{RL}{
  short = RL ,
  long  = reinforcement learning ,
  tag = abbrev
}

\DeclareAcronym{DL}{
  short = DL ,
  long  = deep learning ,
  tag = abbrev
}

\DeclareAcronym{NLP}{
  short = NLP ,
  long  = natural language processing ,
  tag = abbrev
}

\DeclareAcronym{SVM}{
  short = SVM ,
  long  = support vector machine ,
  tag = abbrev
}

\DeclareAcronym{SEI}{
    short = SEI ,
    long = specific emitter identification ,
    tag = abbrev
}

\DeclareAcronym{IQ}{
  short = IQ,
  long  = In-phase/Quadrature ,
  tag = abbrev
}

\DeclareAcronym{SNR}{
    short = SNR ,
    long = signal-to-noise ratio ,
    tag = abbrev
}

\DeclareAcronym{RFML}{
  short = RFML ,
  long  = radio frequency machine learning ,
  tag = abbrev
}

\DeclareAcronym{AWGN}{
  short = AWGN ,
  long  = additive white Gaussian noise ,
  tag = abbrev
}

\DeclareAcronym{DARPA}{
  short = DARPA ,
  long  = Defense Advanced Research Projects Agency ,
  tag = abbrev
}

\DeclareAcronym{IARPA}{
  short = IARPA ,
  long  = Intelligence Advanced Research Projects Activity ,
  tag = abbrev
}

\DeclareAcronym{LEEP}{
  short = LEEP ,
  long  = Log Expected Empirical Prediction ,
  tag = abbrev
}

\DeclareAcronym{LogME}{
  short = LogME ,
  long  = Logarithm of Maximum Evidence ,
  tag = abbrev
}

\DeclareAcronym{RFMLS}{
  short = RFMLS ,
  long  = Radio Frequency Machine Learning Systems ,
  tag = abbrev
}

\DeclareAcronym{SCISRS}{
  short = SCISRS ,
  long  = Securing Compartmented Information with Smart Radio Systems,
  tag = abbrev
}

\DeclareAcronym{DySPAN}{
  short = DySPAN,
  long  = IEEE International Symposium on Dynamic Spectrum Access Networks ,
  tag = abbrev
}

\DeclareAcronym{GBC}{
  short = GBC,
  long  = Gaussian Bhattacharyya Coefficient ,
  tag = abbrev
}

\DeclareAcronym{NCE}{
  short = NCE,
  long  = Negative Conditional Entropy ,
  tag = abbrev
}

\DeclareAcronym{JC-NCE}{
  short = JC-NCE,
  long  = Joint Correspondences Negative Conditional Entropy ,
  tag = abbrev
}

\DeclareAcronym{OTCE}{
  short = OTCE,
  long  = Optimal Transport-based Conditional Entropy ,
  tag = abbrev
}

\begin{document}

%%
%% The "title" command has an optional parameter,
%% allowing the author to define a "short title" to be used in page headers.
\title{Assessing the Value of Transfer Learning Metrics for RF Domain Adaptation}

%%
%% The "author" command and its associated commands are used to define
%% the authors and their affiliations.
%% Of note is the shared affiliation of the first two authors, and the
%% "authornote" and "authornotemark" commands
%% used to denote shared contribution to the research.
\author{Lauren J. Wong}
\email{ljwong@vt.edu}
\orcid{1234-5678-9012}
\affiliation{%
  \institution{National Security Institute, Virginia Tech}
  \city{Blacksburg}
  \state{Virginia}
  \country{USA}
}
\affiliation{%
  \institution{Intel AI Lab}
  \city{Santa Clara}
  \state{California}
  \country{USA}
}

\author{Sean McPherson}
\email{sean.mcpherson@intel.com}
\orcid{1234-5678-9012}
\affiliation{%
  \institution{Intel AI Lab}
  \city{Santa Clara}
  \state{California}
  \country{USA}
}

\author{Alan J. Michaels}
\email{ajm@vt.edu}
\orcid{1234-5678-9012}
\affiliation{%
  \institution{National Security Institute, Virginia Tech}
  \city{Blacksburg}
  \state{Virginia}
  \country{USA}
}

%%
%% By default, the full list of authors will be used in the page
%% headers. Often, this list is too long, and will overlap
%% other information printed in the page headers. This command allows
%% the author to define a more concise list
%% of authors' names for this purpose.
\renewcommand{\shortauthors}{Wong, et al.}

%%
%% The abstract is a short summary of the work to be presented in the
%% article.
\begin{abstract}
The use of \ac{TL} techniques has become common practice in fields such as \ac{CV} and \ac{NLP}.
Leveraging prior knowledge gained from data with different distributions, \ac{TL} offers higher performance and reduced training time, but has yet to be fully utilized in applications of \ac{ML} and \ac{DL} techniques to applications related to wireless communications, a field loosely termed \ac{RFML}.
This work begins this examination by evaluating the how \ac{RF} domain changes encourage or prevent the transfer of features learned by \ac{CNN}-based automatic modulation classifiers. 
Additionally, we examine existing \textit{transferability} metrics, \ac{LEEP} and \ac{LogME}, as a means to both select source models for \ac{RF} domain adaptation and predict post-transfer accuracy without further training.
% Additionally, we show existing \textit{transferability} metrics, \ac{LEEP} and \ac{LogME}, strongly correlate with post-transfer accuracy, are suitable for use in \ac{RFML}, and present an algorithm for using such transferability metrics to predict post-transfer accuracy without further training.
\end{abstract}

%%
%% The code below is generated by the tool at http://dl.acm.org/ccs.cfm.
%% Please copy and paste the code instead of the example below.
%%


%%
%% Keywords. The author(s) should pick words that accurately describe
%% the work being presented. Separate the keywords with commas.

%%
%% This command processes the author and affiliation and title
%% information and builds the first part of the formatted document.
\maketitle

\section{Introduction}
Following the release of the \ac{DARPA} \ac{RFMLS} program in 2017 \cite{rfmls}, research into the application of \ac{ML} and \ac{DL} techniques to wireless communications problems has risen significantly. 
The field of \ac{RFML} continues to be the topic of many government research programs and academic conferences including the \ac{IARPA} \ac{SCISRS} program \cite{scisrs} and the \ac{DySPAN} \cite{dyspan2021}.
As a result, advances in \ac{RFML} have yielded increased performance and flexibility in applications such as spectrum awareness, cognitive radio, and networking while reducing the need for expert-defined pre-processing and feature extraction techniques  \cite{west2017}.

\begin{figure}[t]
    \centering
    \begin{subfigure}[b]{.45\textwidth}
      \centering
      \includegraphics[width=0.9\linewidth]{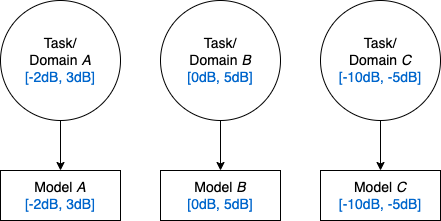}
      \caption{Traditional Machine Learning}
      \label{fig:tl_stock1}
    %   \vspace{0.25cm}
    \end{subfigure}
    % \hfill
    \begin{subfigure}[b]{.45\textwidth}
      \centering
      \includegraphics[width=0.9\linewidth]{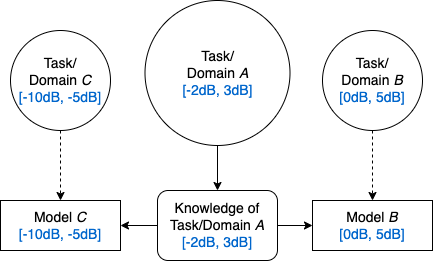}
      \caption{Transfer Learning}
      \label{fig:tl_stock2}
    \end{subfigure}
    \caption{In traditional \ac{ML} (Fig. \ref{fig:tl_stock1}), a new model is trained from random initialization for each domain/task pairing. \ac{TL} (Fig. \ref{fig:tl_stock2}) utilizes prior knowledge learned on one domain/task, in the form of a pre-trained model, to improve performance on a second domain and/or task. A concrete example for environmental adaptation to \ac{SNR} is given in blue.}
    \label{fig:tl_stock}
    % \vspace{-0.25cm}
\end{figure}

Current state-of-the-art \ac{RFML} techniques rely upon supervised learning techniques trained from random initialization, and thereby assume the availability of a large corpus of labeled training data (synthetic, captured, or augmented \cite{clark2020}), which is representative of the anticipated deployed environment.
Over time, this assumption inevitably breaks down as a result of changing hardware and channel conditions, and as a consequence, performance degrades significantly \cite{hauser2018, sankhe2019}.
\Ac{TL} techniques can be used to mitigate these performance degradations by using prior knowledge obtained from a \textit{source} domain and task, in the form of learned representations, to improve performance on a ``similar" \textit{target} domain and task using less data, as depicted in Fig. \ref{fig:tl_stock}.

\begin{figure*}
    \centering
    \includegraphics[width=0.9\textwidth]{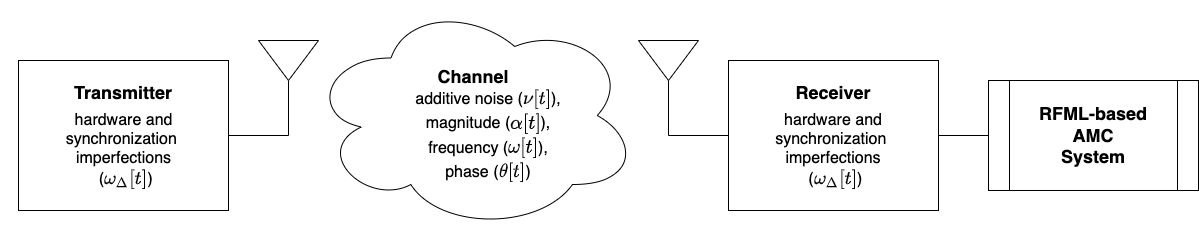}
    \caption{A system overview of the \ac{RF} hardware and channel environment simulated in this work with the parameters/variables ($\alpha[t]$, $\omega[t]$, $\theta[t]$, $\nu[t]$, $\omega_\Delta[t]$) that each component of the system has the most significant impact on.}
    \label{fig:system}
\end{figure*}

Though \ac{TL} techniques have demonstrated significant benefits in fields such as \ac{CV} and \ac{NLP} \cite{ruder2019}, including higher performing models, significantly less training time, and far fewer training samples \cite{olivas2009}, \cite{wong2022} showed that the use of \ac{TL} in \ac{RFML} is currently lacking through the construction of an \ac{RFML} specific \ac{TL} taxonomy.
% The limited works surveyed in \textcolor{red}{[CITE taxonomy]} showed sequential \ac{TL} techniques, where the source domain/task is learned first and the target domain/task is learned during a second training phase (i.e. a head re-training or fine-tuning stage), provide an avenue for increased performance over a wider variety of hardware platforms and channel conditions with reduced amounts of captured training data \cite{kuzdeba2021}.
This work addresses current limitations in understanding how the training domain in particular, characterized by the \ac{RF} hardware and the channel environment \cite{wong2022} depicted in Fig. \ref{fig:system}, impacts learned behavior and therefore facilitates or prevents successful transfer.

% For example, while \ac{CV} algorithm performance is not significantly impacted by a change in the camera(s) used to collect data, so long as the image resolution remains consistent \cite{liu2020}, work in \cite{hauser2018} showed that a change in transmitter/receiver pairs negatively impacted performance by as much as 7\%, despite the collection parameters and even the brand/models of transmitters/receivers remaining consistent.
% Therefore, \textit{platform adaptation} techniques that transfer knowledge gleaned from one hardware platform (or set of platforms) to a second hardware platform (or set of platforms) are a necessity in \ac{RFML}, but not in \ac{CV}.

The contributions of this work are three fold:
First, this work systematically evaluates \ac{RF} domain adaptation performance as a function of several parameters of interest for an \ac{AMC} use-case \cite{hauser2018}:
\begin{easylist}[itemize]
@ \Ac{SNR}, which represents a change in the \ac{RF} channel environment (i.e., an increase/decrease in the additive interference, $\nu[t]$, of the channel) and/or transmitting devices (i.e., an increases/decrease in the magnitude, $\alpha[t]$, of the transmitted signal) and an \textit{environment adaptation} problem, %The source and target domains differ.
@ \Ac{FO}, which represents a change in the transmitting and/or receiving devices (i.e., an increase/decrease in $\omega_\Delta[t]$ due to hardware imperfections or a lack of synchronization) and a \textit{platform adaptation} problem, and %The source and target domains differ.
@ Both \ac{SNR} and \ac{FO}, representing a change in both the \ac{RF} channel environment \emph{and} the transmitting/receiving devices and an \textit{environment platform co-adaptation} problem.
\end{easylist}
\noindent These three parameter sweeps address each type of \ac{RF} domain adaptation discussed in the \ac{RFML} \ac{TL} taxonomy \cite{wong2022}, and resulted in the construction of 82 training sets, 82 validation sets, and 82 test sets and the training and evaluation of 4442 models.
From these experiments, we identify a number of practical takeaways for how best to perform \ac{RF} domain adaptation including how changes in \ac{SNR} and \ac{FO} impact the difficulty of \ac{AMC}, an intuitive notion of source/target domain similarity induced by these parameter sweeps, and the impact these have on transfer performance, as well as a comparison of head re-training versus fine-tuning for \ac{RF} \ac{TL}.

The second contribution of this work regards how transferability is measured herein.
Intuitively, post-transfer top-1 accuracy provides the ground truth measure of transferability, as a measure of performance after transfer learning has occurred through head re-training or fine-tuning of the source model.
However, in the scenario where many source models are available for transfer to an alternate domain, evaluating the post-transfer top-1 accuracy for each source model may be too time consuming and computationally expensive.
Therefore, in addition to evaluating the transferability of pre-trained source models using post-transfer top-1 accuracy, we examine two existing transferability metrics not known to be used in the \ac{RF} domain prior to this work: \ac{LEEP} \cite{nguyen2020} and \ac{LogME} \cite{you2021}.
Transferability metrics, such as \ac{LEEP} and \ac{LogME}, provide a measure of how well a source model will transfer to a target dataset, and are evaluated without performing transfer learning, using only a single forward pass through the source model.
Though \ac{LEEP} and \ac{LogME} are designed to be modality independent, we confirm that they are suitable for use in \ac{RFML} by showing that \ac{LEEP} and \ac{LogME} strongly correlate with post-transfer top-1 accuracy, as well as with each other.

Third, we present a method for using transferability metrics such as these to predict post-transfer accuracy, within a confidence interval, and without further training.
More specifically, given a labelled target raw \ac{IQ} dataset and a selection of pre-trained source models, we show that transferability metrics such as \ac{LEEP} and/or \ac{LogME} can be used to provide a lower and upper bound on how well each source model will perform once transferred to the target dataset, without performing head re-training or fine-tuning.

%%%%%%%%%%%%%%%%%%%%%%%%%%%%%%%%%%%%%%%%%%%%
In total, this work addresses a number of key research questions, including:
\begin{easylist}[enumerate]
@ When and how is \ac{RF} domain adaptation most successful? -- Through exhaustive experimentation, this work provides generalized and practical guidelines for using \ac{TL} for \ac{RF} domain adaptation that are intuitive and consistent with the general theory of \ac{TL} and domain adaptation.
@ Can transferability metrics, such as \ac{LEEP} and \ac{LogME}, be used to effectively select models for \ac{RF} domain adaptation? -- This work shows that both \ac{LEEP} and \ac{LogME} strongly correlate with post-transfer top-1 accuracy in the context of this \ac{AMC} use-case, and that results are consistent with those shown in the original publications. 
@ Can transferability metrics, such as \ac{LEEP} and \ac{LogME}, predict post-transfer accuracy? -- This work presents such an approach, with no additional re-training requirements and confidence intervals provided.
\end{easylist}

This paper is organized as follows:
Section \ref{sec:related} provides requisite background knowledge, and discusses related and prior works in \ac{TL} for \ac{RFML}, as well as transferability metrics and transfer accuracy prediction in other modalities such as \ac{CV} and \ac{NLP}.
In Section \ref{sec:methodology}, each of the key methods and systems used and developed for this work are described in detail, including the simulation environment and dataset creation, the model architecture and training, and the transferability metrics.
Section \ref{sec:results} presents experimental results and analysis, addressing the key research questions described above, and the proposed post-transfer accuracy prediction method.
Section \ref{sec:fw} highlights several directions for future work including extensions of this work performed herein using alternative transferability metrics and captured and/or augmented data, generalizations of this work to inductive \ac{TL} settings, and developing more robust or \ac{RF}-specific transferability metrics.
Finally, Section \ref{sec:conclusion} offers conclusions about the effectiveness of \ac{TL} and existing transferability metrics for \ac{RFML} and next steps for incorporating and extending \ac{TL} techniques in \ac{RFML}-based research.
A list of the acronyms used in this work is provided in the appendix for reference.

\section{Background \& Related Work}\label{sec:related}

The recent \ac{RFML} and \ac{TL} taxonomies and surveys  \cite{wong2022, wong2021ecosystem} highlight the limited existing works that successfully use sequential \ac{TL} techniques for both domain adaptation and inductive transfer including transferring pre-trained models across channel environments \cite{chen2019, pati2020}, across wireless protocols \cite{kuzdeba2021, robinson2021}, and from synthetic data to real data \cite{oshea2018, dorner2018, zheng2020, cyborg}, as well as to add or remove output classes \cite{peng2020}, for tasks such as signal detection, \ac{AMC}, and \ac{SEI}. 
However, outside of observing a lack of direct transfer \cite{hauser2018, clark2020, merchant2019}, little-to-no work has examined what characteristics within \ac{RF} data facilitate or restrict transfer \cite{wong2022}.
Without such knowledge, \ac{TL} algorithms for \ac{RFML} are generally restricted to those borrowed from other modalities, such as \ac{CV} and \ac{NLP}.
While correlations can be drawn between the vision or language spaces and the \ac{RF} space, these parallels do not always align, and therefore algorithms designed for \ac{CV} and \ac{NLP} may not always be appropriate for use in \ac{RFML}.
In this work, post-transfer top-1 accuracy is paired with existing transferability metrics, \ac{LEEP} and \ac{LogME}, to identify how changes in the \ac{RF} domain impact transferability, and how \ac{LEEP} and \ac{LogME} can be of value when performing \ac{RF} domain adaptation in practice, namely for model selection and post-transfer accuracy prediction.

\subsection{Transferability Metrics}
As discussed previously, \ac{TL} techniques use prior knowledge obtained from a \textit{source} domain/task to improve performance on a similar \textit{target} domain/task.
More specifically, \ac{TL} techniques aim to further refine a pre-trained source model using a target dataset and specialized training techniques.
However, not all pre-trained source models will transfer well to a given target dataset.
Though it is generally understood that \ac{TL} is successful when the source and target domains/tasks are ``similar" \cite{pan2010}, this notion of source/target similarity is ill-defined.
The goal of a transferability metric is to quantify how well a given pre-trained source model will transfer to a target dataset.
While the area of transferability metrics is growing increasingly popular, to our knowledge, no prior works have examined these metrics in the context of \ac{RFML}.
Transferability metrics developed and examined in the context of other modalities can broadly be categorized in one of two ways: those requiring partial re-training and those that do not.

Partial re-training methods such as Taskonomy \cite{zamir2018} and Task2Vec \cite{achille2019} require some amount of training to occur, whether that be the initial stages of \ac{TL}, full \ac{TL}, or the training of an additional \textit{probe} network, in order to quantify transferability.
Partial re-training methods are typically used to identify relationships between source and target tasks and are useful in meta-learning settings, but are not well suited to settings where time and/or computational resources are limited.
Though the computational complexity of partial re-training methods varies, it vastly exceeds the computational complexity of methods that do not require any additional training, such as those used in this work.

This work focuses on methods that do not require additional training, which typically use a single forward pass through a pre-trained model to ascertain transferability.
Methods such as these are often used to select a pre-trained model from a model library for transfer to a target dataset, a problem known as \textit{source model selection}.
\ac{LEEP} \cite{nguyen2020} and \ac{LogME} \cite{you2021} were chosen for this work having outperformed similar metrics such as \ac{NCE} \cite{tran2019} and H-scores \cite{bao2019} in \ac{CV} and \ac{NLP}-based experiments, and for their modality agnostic design.
However, several new transferability metrics developed concurrently with this work also show promise including \ac{OTCE} \cite{tan2021a} and \ac{JC-NCE} \cite{tan2021b}, TransRate \cite{huang2021}, and \ac{GBC} \cite{pandy2021}, and may be examined as follow on work.

Related works examine source model ranking or selection procedures \cite{renggli2020, li2021}, which either rank a set of models by transferability or select the model(s) most likely to provide successful transfer.
However, source model ranking or selection methods are less flexible than transferability metrics in online or active learning scenarios.
More specifically, source model ranking or selection methods are unable to identify how a new source model compares to the already ranked/selected source models without performing the ranking/selection procedure again.
Related works also include methods for selecting the best data to use for pre-training \cite{bhattacharjee2020} or during the transfer phase \cite{ruder2017}, and approaches to measuring domain, task, and/or dataset similarity \cite{kashyap2020}.

\subsection{Predicting Transfer Accuracy}
The problem of predicting transfer accuracy is still an open field.
To the best of our knowledge, no prior works have examined predicting transfer accuracy specifically for \ac{RFML}, but approaches have been developed for other modalities.
Most similar to our work is the approach given in \cite{van2010}, where the authors showed linear correlation between several domain similarity metrics and transfer accuracy, using statistical inference to derive performance predictions for \ac{NLP} tools.
Similarly, work in \cite{elsahar2019} used domain similarity metrics to predict performance drops as a result in domain shift. 
More recently, \cite{pogrebnyakov2021} proposed using a simple \ac{MLP} to determine how well a source dataset will transfer to a target dataset, again in an \ac{NLP} setting.

\begin{table*}[]
\centering
\caption{Signal types included in this work and generation parameters.}
\label{tab:signals}
% \begin{minipage}{0.5\textwidth}
\small{
\begin{tabular}[t]{@{}ll@{}}
\toprule
\multicolumn{1}{c}{\begin{tabular}[c]{@{}c@{}}Modulation\\ Name\end{tabular}} &  \multicolumn{1}{c}{\begin{tabular}[c]{@{}c@{}}Parameter\\ Space\end{tabular}} \\ \midrule 
BPSK & \begin{tabular}[c]{@{}l@{}}Symbol Order \{2\}\\ RRC Pulse Shape\\ Excess Bandwidth \{0.35, 0.5\}\\ Symbol Overlap $\in$ {[}3, 5{]}\end{tabular} \\ \\
QPSK & \begin{tabular}[c]{@{}l@{}}Symbol Order \{4\}\\ RRC Pulse Shape\\ Excess Bandwidth \{0.35, 0.5\}\\ Symbol Overlap $\in$ {[}3, 5{]}\end{tabular} \\ \\
PSK8  & \begin{tabular}[c]{@{}l@{}}Symbol Order \{8\}\\ RRC Pulse Shape\\ Excess Bandwidth \{0.35, 0.5\}\\ Symbol Overlap $\in$ {[}3, 5{]}\end{tabular} \\ \\
PSK16 & \begin{tabular}[c]{@{}l@{}}Symbol Order \{16\}\\ RRC Pulse Shape\\ Excess Bandwidth \{0.35, 0.5\}\\ Symbol Overlap $\in$ {[}3, 5{]}\end{tabular} \\ \\
OQPSK & \begin{tabular}[c]{@{}l@{}}Symbol Order \{4\}\\ RRC Pulse Shape\\ Excess Bandwidth \{0.35, 0.5\}\\ Symbol Overlap $\in$ {[}3, 5{]}\end{tabular} \\ \\
QAM16 & \begin{tabular}[c]{@{}l@{}}Symbol Order \{16\}\\ RRC Pulse Shape\\ Excess Bandwidth \{0.35, 0.5\}\\ Symbol Overlap $\in$ {[}3, 5{]}\end{tabular} \\ \\
QAM32 & \begin{tabular}[c]{@{}l@{}}Symbol Order \{32\}\\ RRC Pulse Shape\\ Excess Bandwidth \{0.35, 0.5\}\\ Symbol Overlap $\in$ {[}3, 5{]}\end{tabular} \\ \\
QAM64 & \begin{tabular}[c]{@{}l@{}}Symbol Order \{64\}\\ RRC Pulse Shape\\ Excess Bandwidth \{0.35, 0.5\}\\ Symbol Overlap $\in$ {[}3, 5{]}\end{tabular} \\ \\
APSK16 & \begin{tabular}[c]{@{}l@{}}Symbol Order \{16\}\\ RRC Pulse Shape\\ Excess Bandwidth \{0.35, 0.5\}\\ Symbol Overlap $\in$ {[}3, 5{]}\end{tabular} \\ \\ \bottomrule
\end{tabular}
\hspace{0.5cm}
% \end{minipage} \hfill
% \begin{minipage}{0.5\textwidth}
\begin{tabular}[t]{@{}ll@{}}
\toprule
\multicolumn{1}{c}{\begin{tabular}[c]{@{}c@{}}Modulation\\ Name\end{tabular}} & \multicolumn{1}{c}{\begin{tabular}[c]{@{}c@{}}Parameter\\ Space\end{tabular}} \\ \midrule
APSK32 & \begin{tabular}[c]{@{}l@{}}Symbol Order \{32\}\\ RRC Pulse Shape\\ Excess Bandwidth \{0.35, 0.5\}\\ Symbol Overlap $\in$ {[}3, 5{]}\end{tabular} \\ \\
FSK5k & \begin{tabular}[c]{@{}l@{}}Carrier Spacing \{5kHz\}\\ Rect Phase Shape\\ Symbol Overlap \{1\}\end{tabular} \\ \\
FSK75k & \begin{tabular}[c]{@{}l@{}}Carrier Spacing \{75kHz\}\\ Rect Phase Shape\\ Symbol Overlap \{1\}\end{tabular} \\ \\
GFSK5k & \begin{tabular}[c]{@{}l@{}}Carrier Spacing \{5kHz\}\\ Gaussian Phase Shape\\ Symbol Overlap \{2, 3, 4\}\\ Beta $\in$ {[}0.3, 0.5{]}\end{tabular} \\ \\
GFSK75k & \begin{tabular}[c]{@{}l@{}}Carrier Spacing \{75kHz\}\\ Gaussian Phase Shape\\ Symbol Overlap \{2, 3, 4 \}\\ Beta $\in$ {[}0.3, 0.5{]}\end{tabular} \\ \\
MSK & \begin{tabular}[c]{@{}l@{}}Carrier Spacing \{2.5kHz\}\\ Rect Phase Shape\\ Symbol Overlap \{1\}\end{tabular} \\ \\
GMSK & \begin{tabular}[c]{@{}l@{}}Carrier Spacing \{2.5kHz\}\\ Gaussian Phase Shape\\ Symbol Overlap \{2, 3, 4\}\\ Beta $\in$ {[}0.3, 0.5{]}\end{tabular} \\ \\
FM-NB & Modulation Index $\in$ {[}0.05, 0.4{]} \\ \\
FM-WB & Modulation Index $\in$ {[}0.825, 1.88{]} \\ \\
AM-DSB & Modulation Index $\in$ {[}0.5, 0.9{]} \\ \\
AM-DSBSC & Modulation Index $\in$ {[}0.5, 0.9{]} \\ \\
AM-LSB & Modulation Index $\in$ {[}0.5, 0.9{]} \\ \\
AM-USB &  Modulation Index $\in$ {[}0.5, 0.9{]} \\ \\
AWGN &  \\ \bottomrule
\end{tabular}}
% \end{minipage}
\end{table*}

\section{Methodology}\label{sec:methodology}
\begin{figure*}[t!]
    \centering
    \begin{subfigure}[b]{.9\textwidth}
      \centering
      \includegraphics[width=0.9\textwidth]{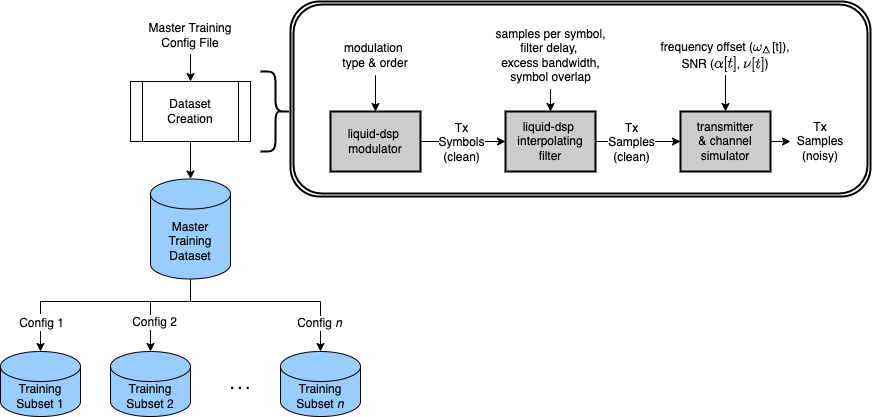}
      \caption{The training dataset generation process, repeated to create the validation and test datasets.}
    %   \vspace{0.5cm}
      \label{fig:sd_data}
    \end{subfigure}
    % \hfill
    \begin{subfigure}[b]{.9\textwidth}
      \centering
      \includegraphics[width=0.7\textwidth]{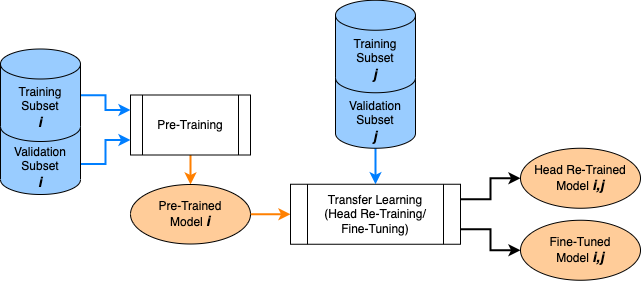}
      \caption{The process for model pre-training and \ac{TL}.}
      \label{fig:sd_training}
    \end{subfigure}
    \begin{subfigure}[b]{.7\textwidth}
      \centering
      \includegraphics[width=0.9\textwidth]{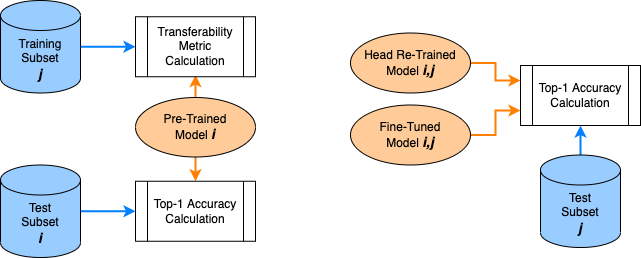}
      \caption{The process for model evaluation and metric calculation.}
      \label{fig:sd_metrics}
    \end{subfigure}
    \caption{A system overview of the (a) dataset creation, (b) model pre-training and \ac{TL}, and (c) model evaluation and transferability metric calculation processes used in this work.}
    \label{fig:system_diagram}
\end{figure*}

This section presents the experimental setup used in this work, shown in Fig. \ref{fig:system_diagram}, which includes the data and dataset creation process, the model architecture and training, and the transferability metrics.
These three key components and processes are each described in detail in the following subsections. 

\subsection{Dataset Creation}\label{sec:data}
This work used a custom synthetic dataset generation tool based off the open-source signal processing library \textit{liquid-dsp} \cite{liquid}, which allowed for full control over the chosen parameters-of-interest, \ac{SNR}, \ac{FO}, and modulation type, and ensured accurate labelling of the training, validation, and test data.
The dataset creation process, shown in Fig. \ref{fig:sd_data}, begins with the construction of a large ``master" dataset containing all combinations of \ac{SNR} and \ac{FO} parameters needed for the experiments performed (Section \ref{sec:master}).
Then, for each of the parameter sweeps performed over \ac{SNR}, \ac{FO}, and both \ac{SNR} and \ac{FO}, subsets of the data were selected from the master dataset using configuration files containing the desired metadata parameters (Sections \ref{sec:snr} - \ref{sec:snr_fo}).
The master dataset created for this work is publicly available on IEEE DataPort \cite{dataport}.

\subsubsection{Simulation Environment}\label{sec:sim_env}
All data used in this work was generated using the same noise generation, signal parameters, and signal types as in \cite{clark2019}.
More specifically, in this work, the signal space has been restricted to the 23 signal types shown in Table \ref{tab:signals}, observed at complex baseband in the form of discrete time-series signals, s[t], where
\begin{equation}
s[t] = \alpha_\Delta[t] \cdot \alpha [t] e^{(j\omega[t]+j\theta[t])} \cdot e^{(j\omega_{\Delta}[t]+j\theta_{\Delta}[t])} + \nu[t] \hspace{0.25cm}
\end{equation}
$\alpha[t]$, $\omega[t]$, and $\theta[t]$ are the magnitude, frequency, and phase of the signal at time $t$, and $\nu[t]$ is the additive interference from the channel.
Any values subscripted with a $\Delta$ represent imperfections/offsets caused by the transmitter/receiver and/or synchronization. 
Without loss of generality, all offsets caused by hardware imperfections or lack of synchronization have been consolidated onto the transmitter during simulation.

Signals are initially synthesized in an \ac{AWGN} channel environment with unit channel gain, no phase offset, and frequency offset held constant for each observation.
Like in \cite{clark2019}, \ac{SNR} is defined as
\begin{equation}
\text{SNR} = 10 \log_{10} \bigg( \frac{\sum_{t=1}^{N-1} \lvert s[t] - \nu[t] \rvert ^2}{\sum_{t=1}^{N-1} \lvert \nu[t] \rvert ^2} \bigg)
\end{equation}
and, with the exception of the \ac{AWGN} signal that has a Nyquist rate of 1, all signals have a Nyquist rate of either 0.5 or 0.33 (twice or three times the Nyquist bandwidth).

\subsubsection{The Master Dataset}\label{sec:master}
The systematic evaluation of transferability as a function of \ac{SNR} and \ac{FO} conducted in this work is possible through the construction of data-subsets with carefully selected metadata parameters from the larger master dataset.
The constructed master dataset contains 600000 examples of each the signal types given in Table \ref{tab:signals}, for a total of 13.8 million examples.
For each example, the \acp{SNR} is selected uniformly at random between [-10dB, 20dB], the \acp{FO} is selected uniformly at random between [-10\%, 10\%] of the sample rate, and all further signal generation parameters such as filtering parameters, symbol order, etc. as specified in Table \ref{tab:signals}. 
Each example and the associated metadata is saved in SigMF format \cite{hilburn2018}.

\begin{figure}[t]
    \centering
    \begin{subfigure}[b]{.45\textwidth}
      \centering
      \includegraphics[width=\linewidth]{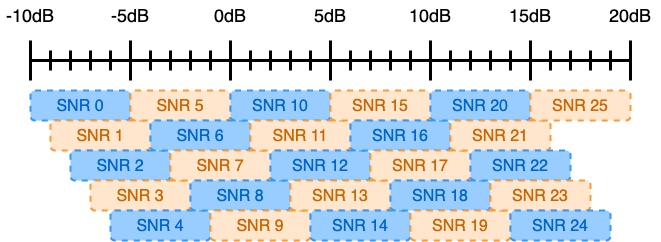}
      \caption{Sweep over SNR.}
      \label{fig:snr_sweep}
      \vspace{0.25cm}
    \end{subfigure}
    % \hfill
    \begin{subfigure}[b]{.45\textwidth}
      \centering
      \includegraphics[width=0.7\linewidth]{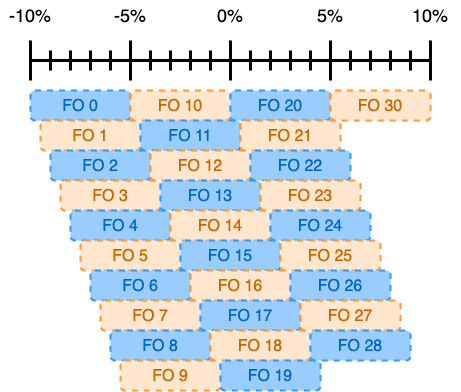}
      \caption{Sweep over FO.}
      \label{fig:fo_sweep}
      \vspace{0.25cm}
    \end{subfigure}
    \begin{subfigure}[b]{.45\textwidth}
      \centering
      \includegraphics[width=\linewidth]{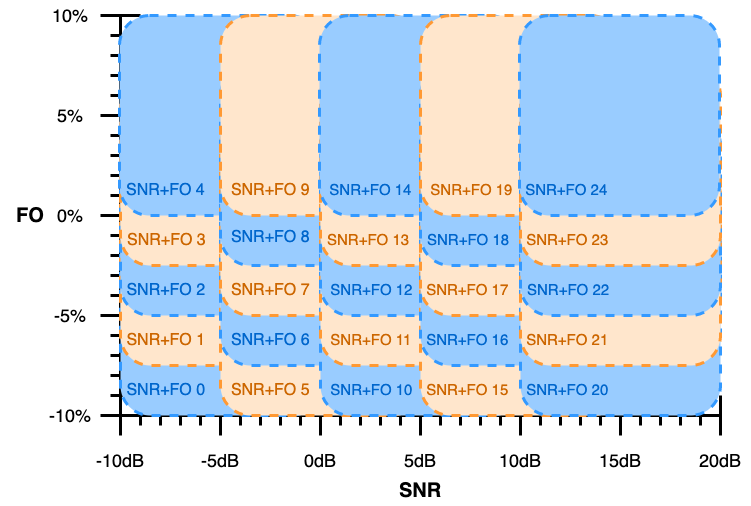}
      \caption{Sweep over SNR and FO.}
      \label{fig:snr_fo_sweep}
    \end{subfigure}
    \caption{The parameter-of-interest range for each data subset  selected from the larger master dataset.}
    \label{fig:data_sweep}
\end{figure}

\subsubsection{\ac{SNR} Sweep}\label{sec:snr}
To analyze the impact of \ac{SNR} alone on transferability, 26 source data-subsets were constructed from the larger master dataset using configuration files, as shown in Fig. \ref{fig:snr_sweep}. 
Each data-subset contains examples with \acp{SNR} selected uniformly at random from a 5dB range sweeping from -10dB to 20dB in 1dB steps (i.e. [-10dB, -5dB], [-9dB, -4dB], ..., [15dB, 20dB]), and for each data-subset in this \ac{SNR} sweep, \ac{FO} was selected uniformly at random between [-5\%, 5\%] of sample rate.
This \ac{SNR} sweep yielded 26 pre-trained source models, each of which was transferred to the remaining 25 target data-subsets (as shown in Fig. \ref{fig:sd_training}), yielding 650 models transferred using head re-training and 650 models transferred using fine-tuning.

\subsubsection{\ac{FO} Sweep}\label{sec:fo}
To analyze the impact of \ac{FO} alone on transferability, 31 source data-subsets were constructed from the larger master dataset (as shown in Fig. \ref{fig:fo_sweep}) containing examples with \acp{FO} selected uniformly at random from a 5\% range sweeping from -10\% of sample rate to 10\% of sample rate in 0.5\% steps (i.e. [-10\%, -5\%], [-9.5\%, -4.5\%], ..., [5\%, 10\%]).
For each data-subset in this \ac{FO} sweep, \ac{SNR} was selected uniformly at random between [0dB, 20dB].
This \ac{FO} sweep yielded 31 pre-trained source models, each of which was transferred to the remaining 30 target data-subsets (as shown in Fig. \ref{fig:sd_training}) yielding 930 models transferred using head re-training, and 930 models transferred using fine-tuning.

\subsubsection{\ac{SNR} $+$ \ac{FO} Sweep}\label{sec:snr_fo}
To analyze the impact of both \ac{SNR} and \ac{FO} on transferability, 25 source data-subsets were constructed from the larger master dataset (as shown in Fig. \ref{fig:snr_fo_sweep}) containing examples with \acp{SNR} selected uniformly at random from a 10dB range sweeping from -10dB to 20dB in 5dB steps (i.e. [-10dB, 0dB], [-5dB, 5dB], ..., [10dB, 20dB]) and with \acp{FO} selected uniformly at random from a 10\% range sweeping from -10\% of sample rate to 10\% of sample rate in 2.5\% steps (i.e. [-10\%, 0\%], [-7.5\%, 2.5\%], ..., [0\%, 10\%]).
This \ac{SNR} and \ac{FO} sweep yielded 25 pre-trained source models, each of which was transferred to the remaining 24 target data-subsets (as shown in Fig. \ref{fig:sd_training}) yielding 600 models transferred using head re-training, and 600 models transferred using fine-tuning.

\subsection{Model Architecture and Training}
% For the purpose of this work, the choice of architecture is largely irrelevant, so long as the architecture is effective and is consistent between all models trained. 
The aim of this work is to use the selected metrics to quantify the ability to transfer the features learned by a single architecture trained across pairwise combinations of source/target datasets with varying 
(1) \acp{SNR}, 
(2) \acp{FO}, or
(3) \acp{SNR} and \ac{FO}
in order to identify the impact of these parameters-of-interest on transferability. 
Given the large number of models trained for this work, training time was a primary concern when selecting the model architecture. 
Therefore, this work uses a simple \ac{CNN} architecture, shown in Table \ref{tab:model}, that is based off of the architectures used in \cite{clark2019} and \cite{wong2021cb}.

\begin{table}[]
\centering
% \vspace{0.25cm}
\caption{Model architecture.}
\label{tab:model}
\small{
\begin{tabular}{@{}lcc@{}}
\toprule
Layer Type & Num Kernels/Nodes & Kernel Size \\ \midrule
Input & size = (2, 128) &  \\
Conv2d & 1500 & (1, 7) \\
ReLU &  &  \\
Conv2d & 96 & (2, 7) \\
ReLU &  &  \\
Dropout & rate = 0.5 &  \\
Flatten &  &  \\
Linear & 65 &  \\
Linear & 23 &  \\ \midrule
\multicolumn{3}{l}{\hspace{1cm}Trainable Parameters: 7434243} \\ \bottomrule
\end{tabular}}
\end{table}

The model pre-training and \ac{TL} process is shown in Fig. \ref{fig:sd_training}, and represents a standard training pipeline. 
For pre-training, the training dataset contained 5000 examples per class, and the validation dataset contained 500 examples per class.
These dataset sizes are consistent with \cite{clark2019} and adequate to achieve consistent convergence. 
Each model was trained using the Adam optimizer \cite{kingma2014} and Cross Entropy Loss \cite{pytorchCE}, with the PyTorch default hyper-parameters \cite{pytorch} (a learning rate of 0.001, without weight decay), for a total of 100 epochs. 
A checkpoint was saved after the epoch with the lowest validation loss, and was reloaded at the conclusion of the 100 epochs.
For both head re-training and model fine-tuning, the training dataset contained 500 examples per class, and the validation dataset contained 50 examples per class, representing a smaller sample of available target data.
The head re-training and fine-tuning processes both used the Adam optimizer and Cross Entropy Loss as well, with checkpoints saved at the lowest validation loss over 100 epochs.
During head re-training, only the final layer of the model was trained, again using the PyTorch default hyper-parameters, while the rest of the model's parameters were frozen.
During fine-tuning, the entire model was trained with a learning rate of 0.0001, an order of magnitude smaller than the PyTorch default of 0.001.

\subsection{Transferability Metrics}\label{sec:metrics}
% \textcolor{red}{Figure \ref{fig:sd_metrics}}
As previously discussed, while transfer accuracy provides the ground truth measure of transferability, calculating transfer accuracy requires performing sequential learning techniques such as head re-training or fine-tuning to completion, in addition to the labelled target dataset.
\ac{LEEP} \cite{nguyen2020} and \ac{LogME} \cite{you2021} are existing metrics designed to predicting how well a pre-trained source model will transfer to a labelled target dataset, without performing transfer learning techniques and using only a single forward pass through the pre-trained source model.
These metrics in particular were shown to outperform similar metrics, \ac{NCE} \cite{tran2019} and H-scores \cite{bao2019}, and are designed to be modality agnostic.
Therefore, though neither metric is known to have been shown to correlate with transfer accuracy in the context of \ac{RFML}, the success both metrics showed in \ac{CV} and \ac{NLP} applications bodes well for the \ac{RF} case.

\acf{LEEP} \cite{nguyen2020} can be described as the ``average log-likelihood of the expected empirical predictor, a simple classifier that makes prediction[s] based on the expected empirical conditional distribution between source and target labels," and has been shown to correlate well with transfer accuracy using image data, even when the target datasets are small or imbalanced.
The metric is bounded between $(-\infty,0]$, such that values closest to zero indicate best transferability, though the scores tend to be smaller when there are more output classes in the target task.
The calculation does not make any assumptions about the similarity of the source/target input data, except that they are the same size.
For example, if the source data is raw \ac{IQ} data of size 2x128, then the target data must also be of size 2x128, but need not be in raw \ac{IQ} format (i.e. the target data could be in polar format).
Therefore, the metric is suitable for estimating transferability when the source and target tasks (output classes) differ.
However, the calculation of the metric does require the use of a Softmax output layer, limiting the technique to supervised classifiers.

\acf{LogME} \cite{you2021} ``estimate[s] the maximum value of label evidence given features extracted by pre-trained models" using a computationally efficient Bayesian algorithm.
More specifically, the pre-trained model is used as a feature extractor, and the \ac{LogME} score is computed using the extracted features and ground truth labels of the target dataset.
Like \ac{LEEP}, the calculation only assumes that the source and target input data are the same size.
The metric is bounded between $[-1, 1]$, such that values closes to -1 indicate worst transferability and values closest to 1 indicate best transferability.
\ac{LogME} does not require the use of a Softmax output layer, and is therefore appropriate in un-supervised settings, regression settings, and the like.
Further, \ac{LogME} was shown to outperform \ac{LEEP} in an image classification setting, better correlating with transfer accuracy, and has also shown positive results in an \ac{NLP} setting.

\section{Experimental Results \& Analysis}\label{sec:results}

The product of the experiments performed herein is 82 data subsets, each with distinct \ac{RF} domains, 82 source models trained from random initialization, and 4360 transfer learned models, half transferred using head re-training and the remaining half transferred using fine-tuning.
Associated with each of the 4360 transfer learned models is a top-1 accuracy value, a \ac{LEEP} score, and a \ac{LogME} score.
Given the careful curation of the signal parameters contained within each data subset, as well as the breadth of signal parameters observed, generalized conclusions can be drawn regarding \ac{TL} performance as function changes in the propagation environment (\ac{SNR}) and transmitter/receiver hardware (\ac{FO}).
However, it should be noted that further experiments using captured data are required in order to draw more concrete guidelines for using \ac{RF} \ac{TL} in the field \cite{clark2020}, and is left for future work.
The following subsections present the results obtained from the experiments performed, and discuss how well \ac{LEEP} and \ac{LogME} perform in the \ac{RF} modality, how to use transferability metrics to predict post-transfer performance, as well as some insights and practical takeaways that can be gleaned from the results given including a preliminary understanding of when and how to use \ac{TL} for \ac{RF} domain adaptation.

%%%%%%%%%%%%%%%%%%%%%%%%%%%%%%%%%%%%%%%%%%%%%%%%%%%%%%
\subsection{When and how is \ac{RF} domain adaptation most successful?}

\begin{figure}[t]
    \centering
    \includegraphics[width=0.75\linewidth]{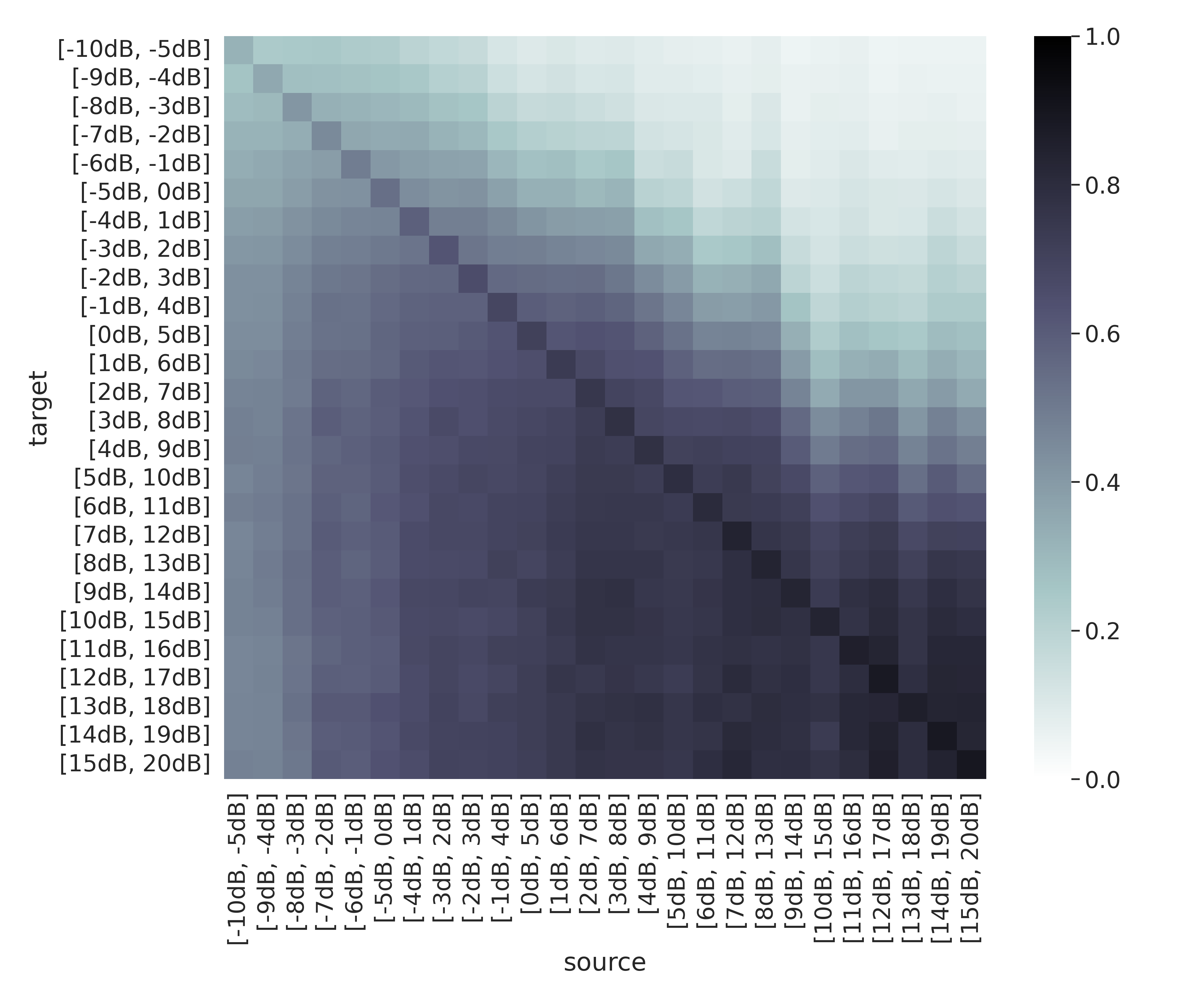}
    \caption{The post-transfer top-1 accuracy for each source/target dataset pair constructed for the sweep over \ac{SNR} using head re-training to perform domain adaptation. When fine-tuning is used to perform domain adaptation, the same trends are apparent.}
    \label{fig:snr_heatmap}
\end{figure}

\begin{figure}[t]
    \centering
    \includegraphics[width=0.75\linewidth]{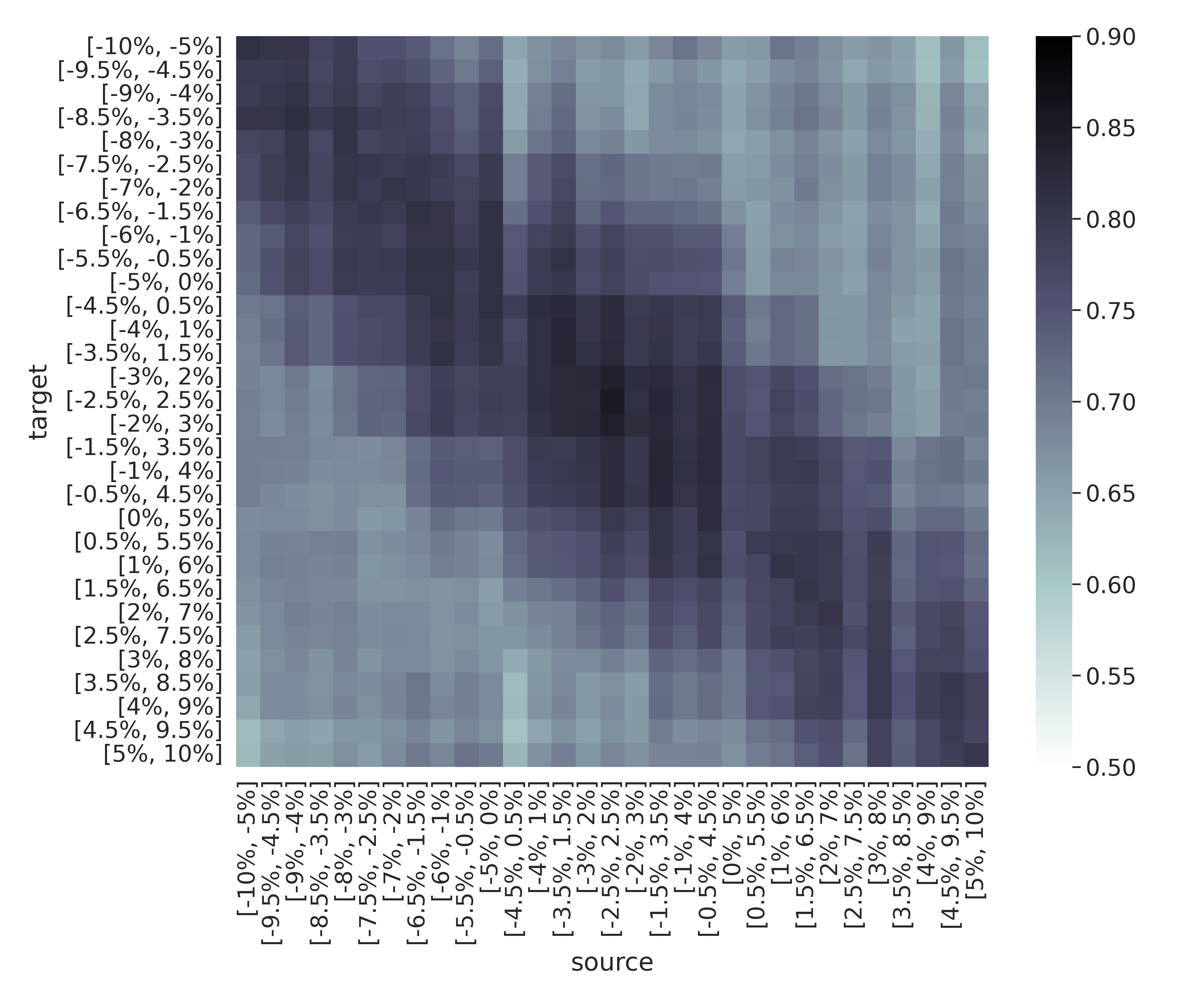}
    \caption{The post-transfer top-1 accuracy for each source/target dataset pair constructed for the sweep over \ac{FO} using head re-training to perform domain adaptation. When fine-tuning is used to perform domain adaptation, the same trends are apparent. Note the scale differs from Figs. \ref{fig:snr_heatmap} and \ref{fig:snr_fo_heatmap}.}
    \label{fig:fo_heatmap}
\end{figure}

\begin{figure}[t]
    \centering
    \includegraphics[width=0.9\linewidth]{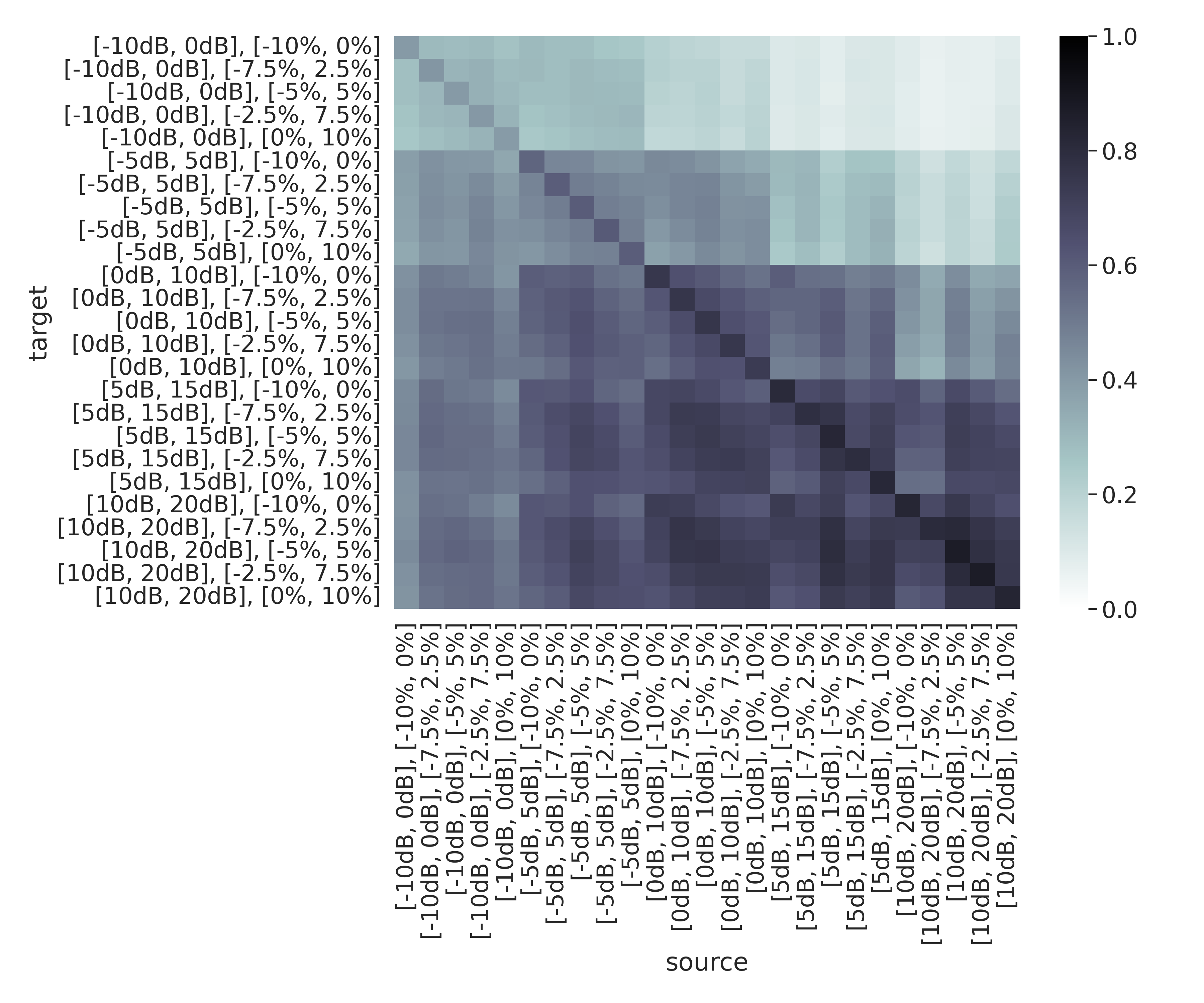}
    \caption{The post-transfer top-1 accuracy for each source/target dataset pair constructed for the sweep over both \ac{SNR} and \ac{FO} using head re-training to perform domain adaptation. When fine-tuning is used to perform domain adaptation, the same trends are apparent.}
    \label{fig:snr_fo_heatmap}
\end{figure}

\subsubsection{Impact of Source/Target Domain Similarity and ``Difficulty" on Transfer Performance}
The heatmaps in Figs. \ref{fig:snr_heatmap}-\ref{fig:snr_fo_heatmap} show the post-transfer top-1 accuracy achieved with each of the source/target pairs.
Note that the post-transfer top-1 accuracy results shown in Figs. \ref{fig:snr_heatmap}-\ref{fig:snr_fo_heatmap} are from the models that used head re-training to transfer from the source to target domains/datasets.
However, the accuracy results from the models that used fine-tuning for transfer show the same trends.

Figs. \ref{fig:snr_heatmap}-\ref{fig:snr_fo_heatmap} show that highest post-transfer performance is achieved along the diagonal of the heatmap, where the source and target domains are most similar.
This behavior is expected, as models trained on similar domains likely learn similar features, and is consistent with the general theory of \ac{TL} \cite{pan2010}, as well as existing works in modalities outside of \ac{RF} \cite{rosenstein2005}.
While the notion of domain similarity is ill-defined in general, for the purposes of this work, we are able to say that domains are more similar when the difference between the source and target \ac{SNR} and/or \ac{FO} ranges is smaller, as all other data generation parameters are held constant. 

Additionally, Figs. \ref{fig:snr_heatmap}-\ref{fig:snr_fo_heatmap} show that transfer across changes in \ac{FO} is approximately symmetric, while transfer across changes in \ac{SNR} are not.
This behavior is also expected, and can be attributed to changes in the relative ``difficulty" between the source and target domains.
More specifically, changing the source/target \ac{SNR} inherently changes the difficulty of the problem, as performing \ac{AMC} in lower \ac{SNR} channel environments is more challenging than performing \ac{AMC} in high \ac{SNR} channel environments.
Therefore, the source models trained on the lower \ac{SNR} ranges will transfer to the higher \ac{SNR} ranges, though may not perform optimally, while the source models trained on the higher \ac{SNR} ranges will fail to transfer to the lower \ac{SNR} ranges, as shown in Figs. \ref{fig:snr_heatmap} and \ref{fig:snr_fo_heatmap}.
In contrast, changing the source/target \ac{FO} does not make performing \ac{AMC} any more or less difficult, but may require modifications to the learned features to accommodate this change.
Moreover, small changes in \ac{FO}, $\omega_\Delta[t]$, in either the positive and negative direction, are expected to perform similarly.
As a result, transfer occurs in either direction off of the diagonal, with best performance closest to the diagonal, as discussed previously.

Practically, these trends indicate that the effectiveness of \ac{RF} domain adaptation increases as the source and target domains become more similar, and, when applicable, \ac{RF} domain adaptation is more often successful when transferring from harder to easier domains.
For example, transferring from [-5dB, 0dB] to [0dB, 5dB] \ac{SNR} is likely more effective than transferring from [5dB, 10dB] to [0dB, 5dB] \ac{SNR}, and transferring from a \ac{FO} range of [-9\%, -4\%] of sample rate to [-8\%, -3\%] of sample rate is likely more effective than transferring from a \ac{FO} range of [-10\%, -5\%] of sample rate to [-8\%, -3\%] of sample rate.

\subsubsection{Environment Adaptation vs. Platform Adaptation}
\begin{figure}[t]
    \centering
    \includegraphics[width=0.55\linewidth]{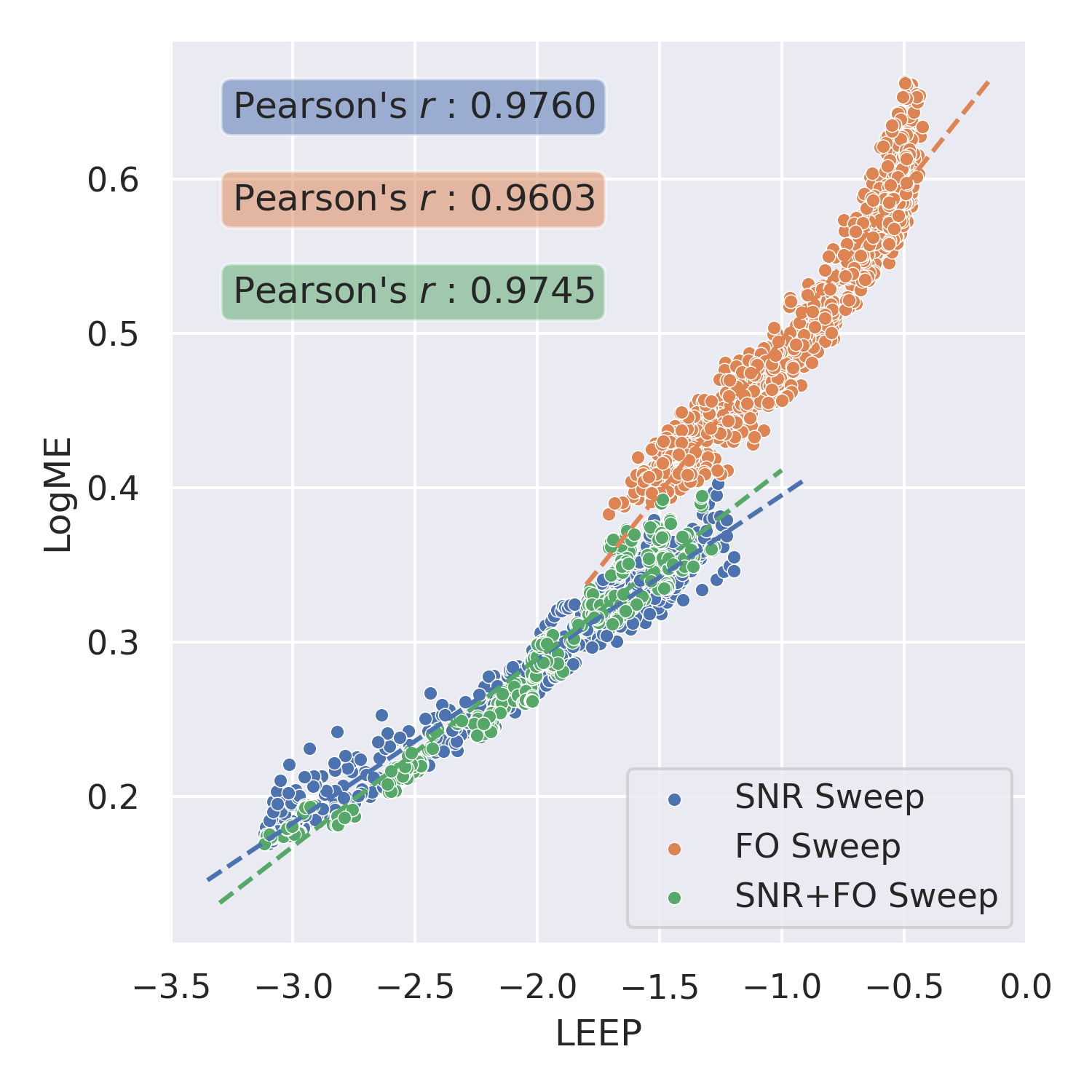}
    \caption{The LEEP versus LogME scores for the sweep over \ac{SNR}, \ac{FO}, and both \ac{SNR} and \ac{FO}. The dashed lines present the linear fit.}
    \label{fig:leepVlogme}
\end{figure}

Recalling that the sweep over \ac{SNR} can be regarded as an environment adaptation experiment and the sweep over \ac{FO} can be regarded as a platform adaptation experiment, more general conclusions can be drawn regarding the challenges that environment and platform adaptation present.
% Given that changes in \ac{SNR} change the relative difficulty of the task where changes in \ac{FO} do not, i
From the discussion in the previous subsection, it follows that changes in \ac{FO} should be easier to overcome than changes in \ac{SNR}.
That is environment adaptation is more difficult to achieve than platform adaptation, and changes in transmitter/receiver hardware are likely easier to overcome using \ac{TL} techniques than changes in the channel environment.
This trend is indirectly shown through the range of accuracies achieved in Figs. \ref{fig:snr_heatmap}-\ref{fig:snr_fo_heatmap}, which is smaller for the \ac{FO} sweep than the \ac{SNR} sweep and \ac{SNR} $+$ \ac{FO} sweep, and is more directly shown in Fig. \ref{fig:leepVlogme}.
(It should be noted that the scale in Fig. \ref{fig:fo_heatmap} differs from Figs. \ref{fig:snr_heatmap} and \ref{fig:snr_fo_heatmap}.)
Fig. \ref{fig:leepVlogme} presents the \ac{LogME} scores as a function of the \ac{LEEP} scores for each of the parameter sweeps performed, showing both the \ac{LEEP} and \ac{LogME} scores are significantly higher for the \ac{FO} sweep than the \ac{SNR} sweep or \ac{SNR} and \ac{FO} sweep indicating better transferability.
(Of course, this conclusion is dependent upon the results presented in Section \ref{sec:metric_acc} which show that \ac{LEEP} and \ac{LogME} correlate with post-transfer accuracy.)
Therefore, in practice, one should consider the similarity of the source/target channel environment before the similarity of the source/target platform, as changes in transmitter/receiver pair are more easily overcome during \ac{TL}.

\subsubsection{Head Re-Training vs. Fine-Tuning}
\begin{figure}[t]
    \centering
    \begin{subfigure}[b]{.49\textwidth}
      \centering
      \includegraphics[width=0.95\linewidth]{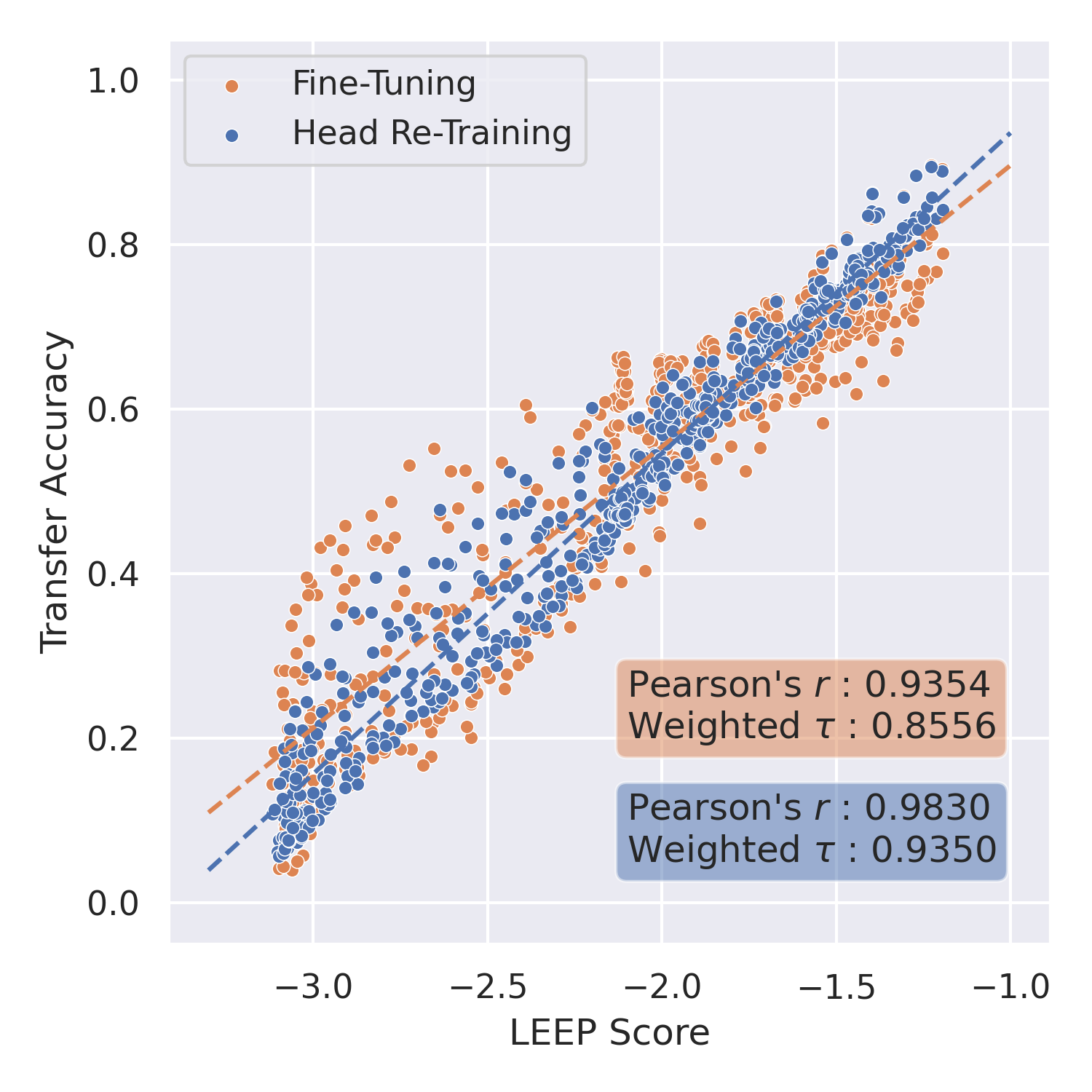}
      \caption{}
      \label{fig:snr_leepVacc}
    \end{subfigure}
    % \hfill
    \begin{subfigure}[b]{.49\textwidth}
      \centering
      \includegraphics[width=0.95\linewidth]{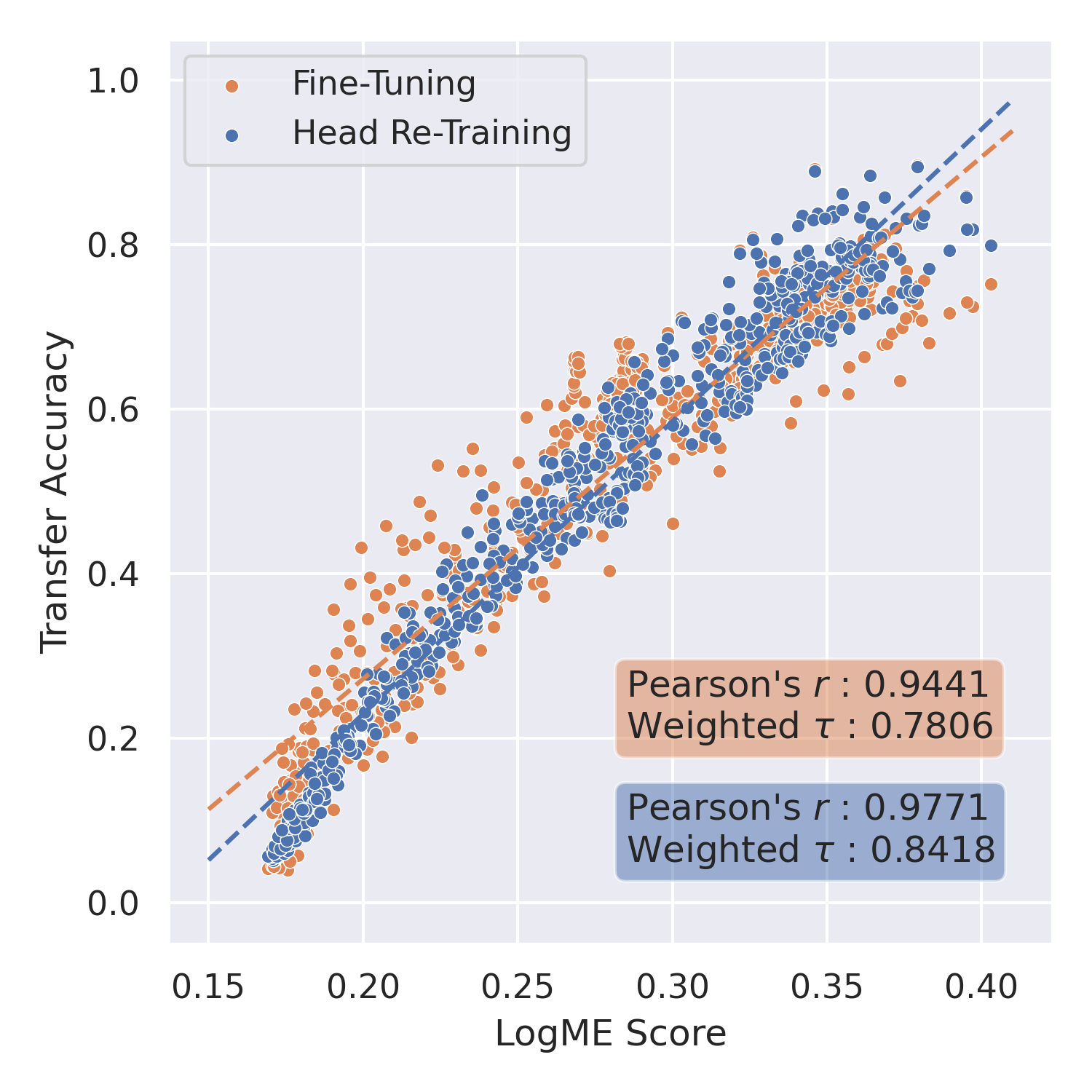}
      \caption{}
      \label{fig:snr_logmeVacc} 
    \end{subfigure}
\caption{The \ac{LEEP} (a) and \ac{LogME} (b) scores versus post-transfer top-1 accuracy for the sweep over \ac{SNR}. The dashed lines present the linear fits for all target domains.}
\label{fig:snr_metricVacc}
\end{figure}

\begin{figure}[t]
    \centering
    \begin{subfigure}[b]{.49\textwidth}
      \centering
      \includegraphics[width=0.95\linewidth]{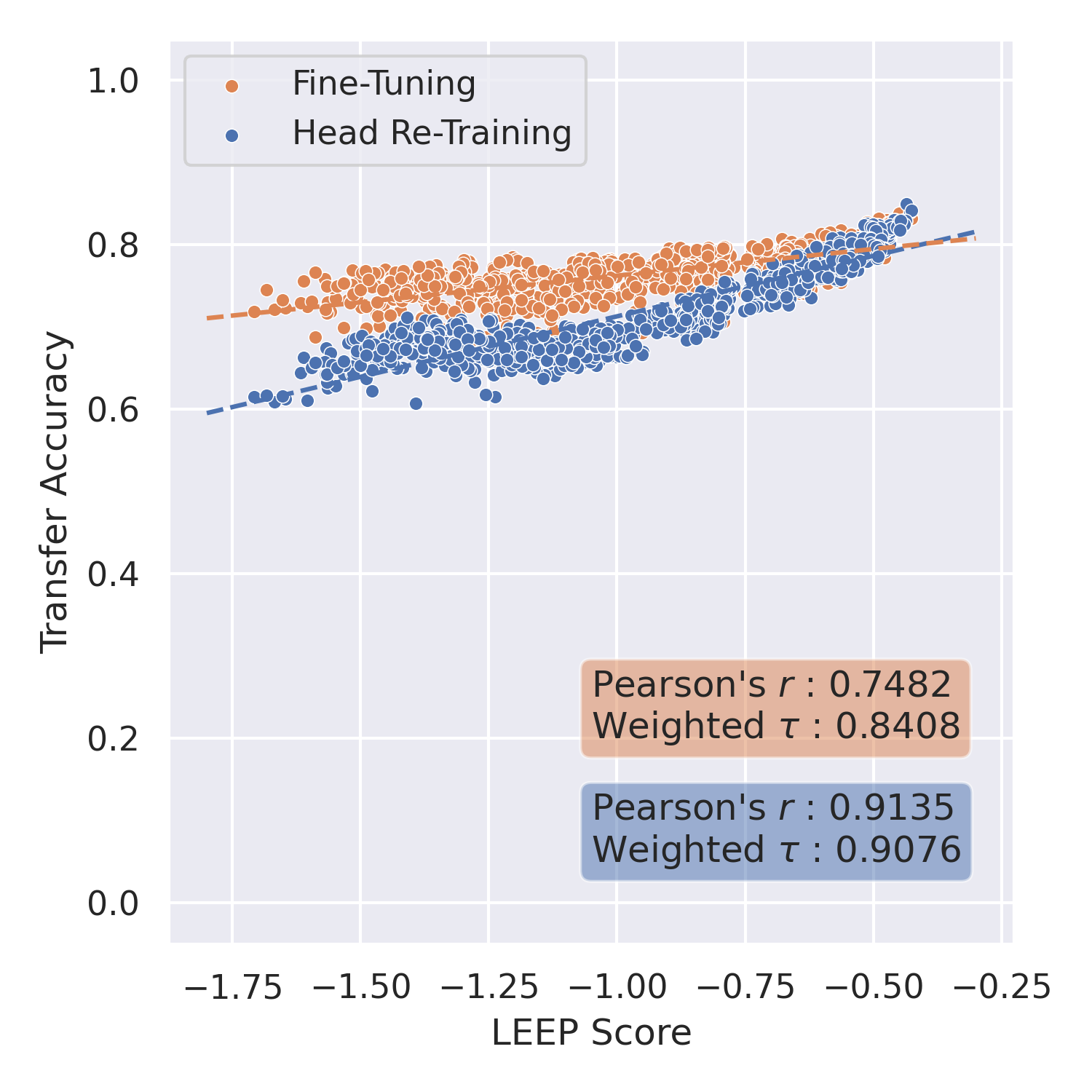}
      \caption{}
      \label{fig:fo_leepVacc}
    \end{subfigure}
    % \hfill
    \begin{subfigure}[b]{.49\textwidth}
      \centering
      \includegraphics[width=0.95\linewidth]{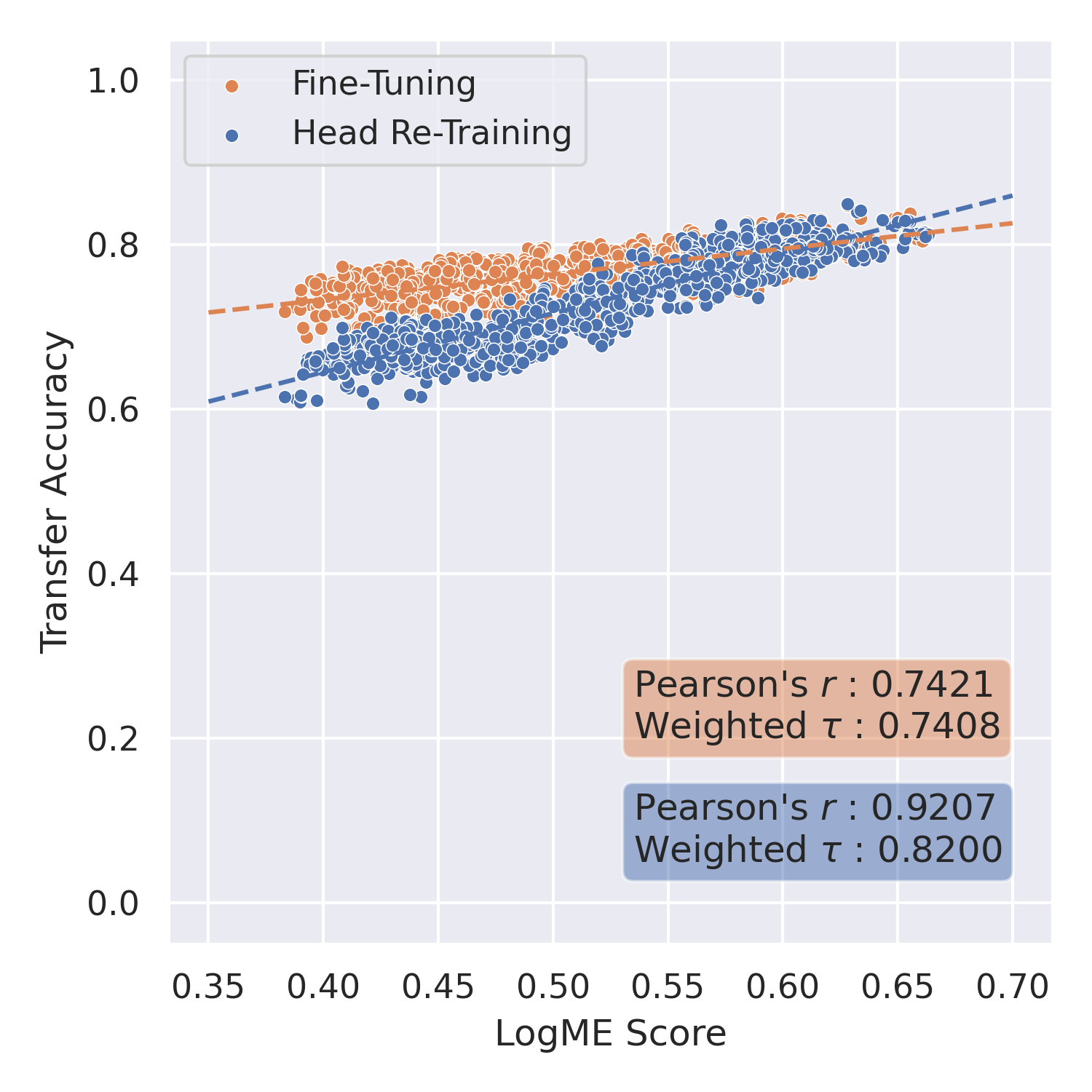}
      \caption{}
      \label{fig:fo_logmeVacc} 
    \end{subfigure}
\caption{The \ac{LEEP} (a) and \ac{LogME} (b) scores versus post-transfer top-1 accuracy for the sweep over \ac{FO}. The dashed lines present the linear fits for all target domains.}
\label{fig:fo_metricVacc}
\end{figure}

\begin{figure}[t]
    \centering
    \begin{subfigure}[b]{.49\textwidth}
      \centering
      \includegraphics[width=0.95\linewidth]{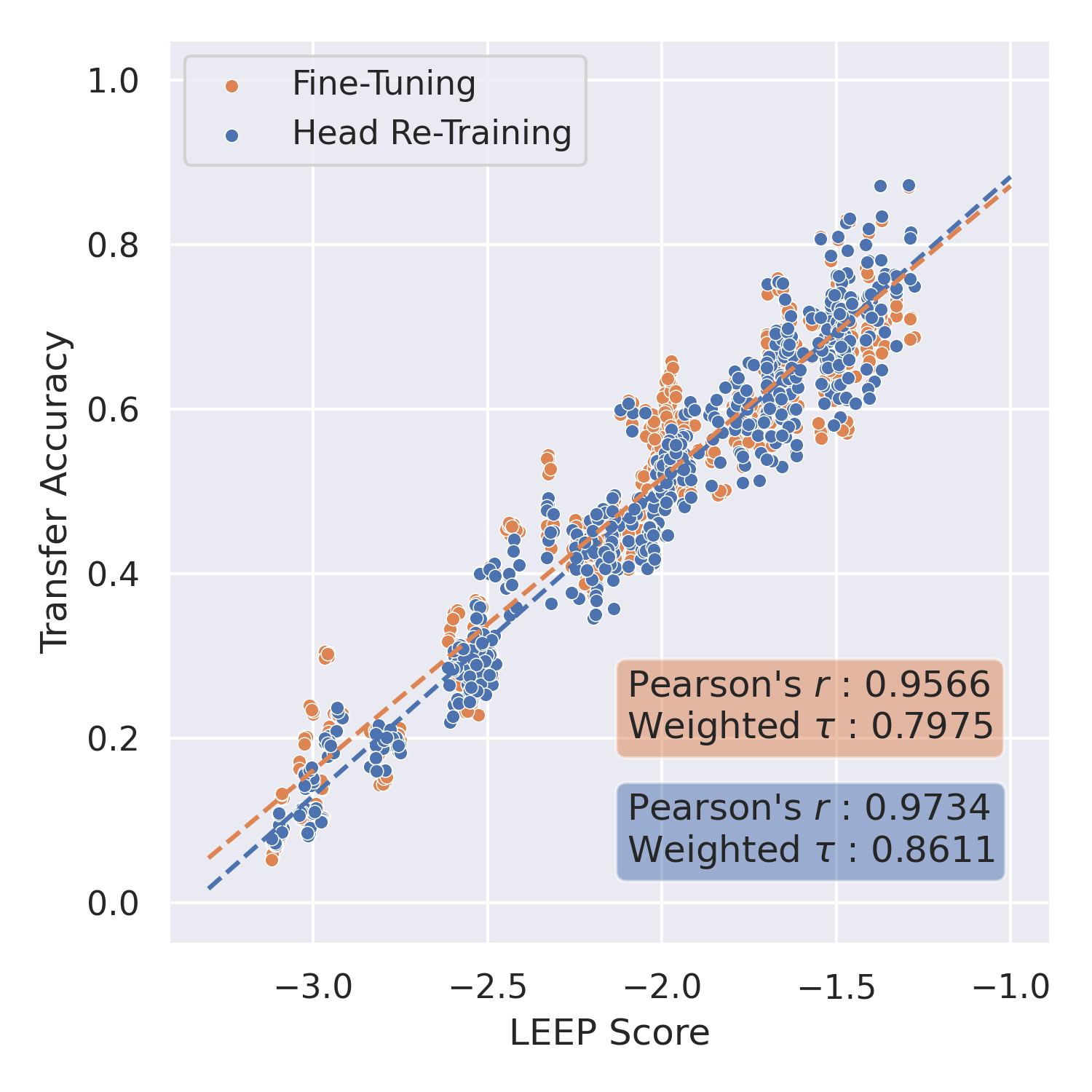}
      \caption{}
      \label{fig:snr_fo_leepVacc}
    \end{subfigure}
    % \hfill
    \begin{subfigure}[b]{.49\textwidth}
      \centering
      \includegraphics[width=0.95\linewidth]{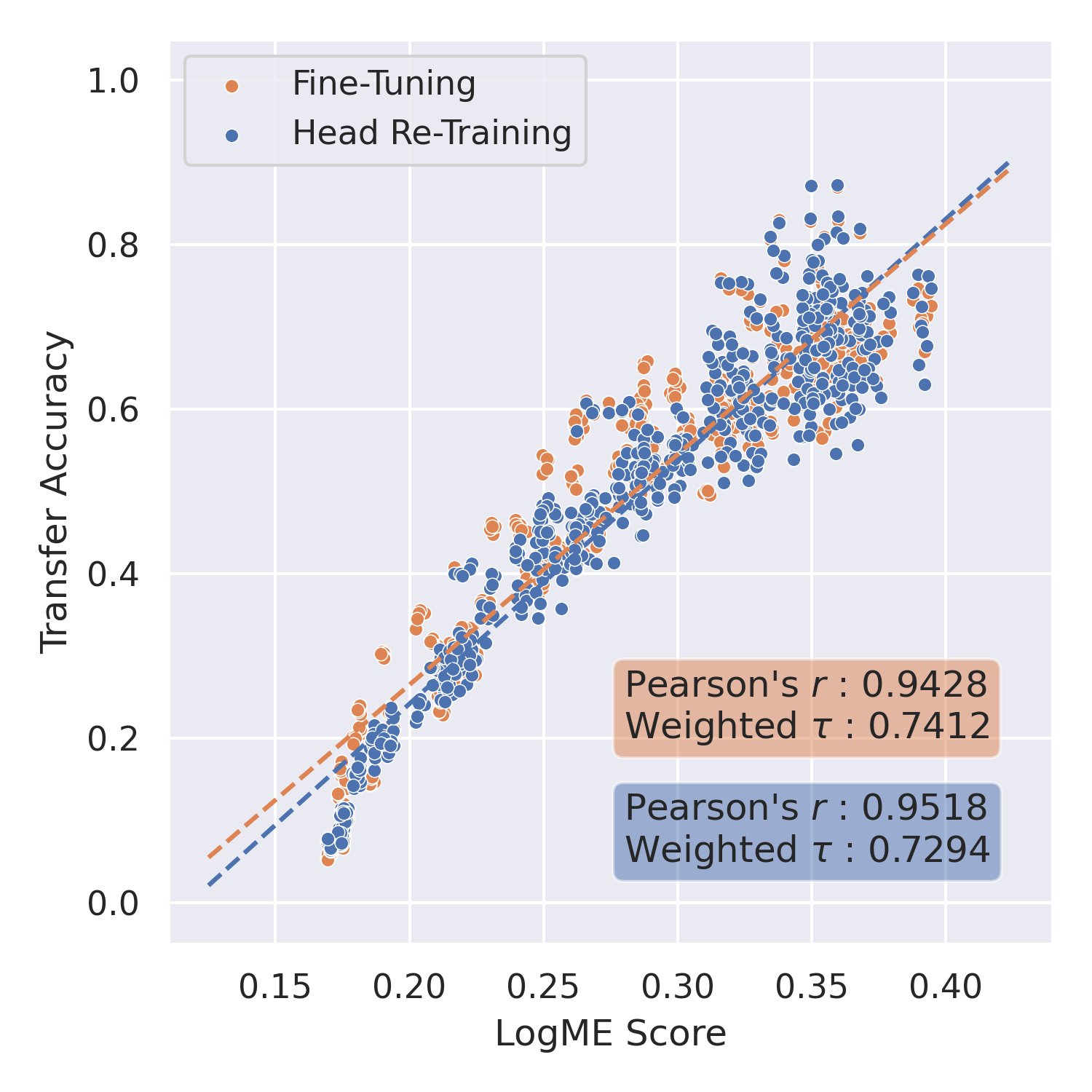}
      \caption{}
      \label{fig:snr_fo_logmeVacc} 
    \end{subfigure}
\caption{The \ac{LEEP} (a) and \ac{LogME} (b) scores versus post-transfer top-1 accuracy for the sweep over both \ac{SNR} and \ac{FO}. The dashed lines present the linear fits for all target domains.}
\label{fig:snr_fo_metricVacc}
\end{figure}

In Figs. \ref{fig:snr_metricVacc}-\ref{fig:snr_fo_metricVacc}, post-transfer top-1 accuracy is shown as a function of either \ac{LEEP} or \ac{LogME}.
These figures indicate that when considering post-transfer top-1 accuracy as the sole performance metric, head re-training is as effective as fine-tuning in \textit{most} settings.
The only setting observed herein in which the fine-tuned models markedly outperformed head re-trained models is in the sweep over \ac{FO}, especially when the \ac{LEEP} and \ac{LogME} scores were low.
A low \ac{LEEP}/\ac{LogME} score indicates a significant change between the source and target domains, in this case a large change in \ac{FO}.
As a result, new features are needed to discern between modulation types, and modifications to the earlier layers of the pre-trained source model, where feature learning occurs, are needed in order to best adapt to the new target domain.

As previously discussed, head re-training is more time efficient and less computationally expensive than fine-tuning, making is a strong case for using head re-training over fine-tuning for \ac{RF} domain adaptation. 
The computational complexity of using head re-training versus fine-tuning is architecture and training algorithm dependent, but as an example, for the \ac{CNN} architecture used in this work and shown in Tab. \ref{tab:model}, the number of trainable parameters for head re-training and fine-tuning is 1,518 and 7,434,243  respectively.
Finally, the correlation coefficients in Figs. \ref{fig:snr_metricVacc}-\ref{fig:snr_fo_metricVacc} show that head re-training is more consistent with \ac{LEEP} and \ac{LogME} scores than fine-tuning, leading to the next discussion of whether \ac{LEEP} and \ac{LogME} scores can be used to select models for \ac{RF} domain adaptation.

%%%%%%%%%%%%%%%%%%%%%%%%%%%%%%%%%%%%%%%%%%%%%%%%%%%%%%%%%%%%%%%%%%%
\subsection{Can transferability metrics, such as \ac{LEEP} and \ac{LogME}, be used to select models for \ac{RF} domain adaptation?}\label{sec:metric_acc}

When evaluating whether a transferability metric is accurate, the primary consideration is how well the metric reflects or correlates with the performance metric(s) used.
Therefore, to identify whether \ac{LEEP} and/or \ac{LogME} can be used to select models for \ac{RF} domain adaptation is to identify how well \ac{LEEP} and \ac{LogME} correlate with post-transfer top-1 accuracy.
To this end, Figs. \ref{fig:snr_metricVacc}-\ref{fig:snr_fo_metricVacc} show \ac{LEEP} and \ac{LogME} versus the achieved transfer accuracy for each of the parameter sweeps described in Section \ref{sec:data}.
These figures qualitatively show that both \ac{LEEP} and \ac{LogME} correlate well with top-1 accuracy after transfer learning, whether through head re-training or fine-tuning for all domain adaptation settings studied.

To quantify whether or not the metrics are useful, two correlation measures are also examined -- the Pearson correlation coefficient \cite{pearsonr} and the weighted $\tau$ \cite{weightedtau} -- specified in the shaded boxes of Figs. \ref{fig:snr_metricVacc}-\ref{fig:snr_fo_metricVacc}.
The Pearson correlation coefficient, or Pearson's $r$, is a measure of linear correlation between two variables used in a wide variety of works, including the original \ac{LEEP} paper.
However, Pearson's $r$ makes a number of assumptions about the data, some of which may not be met by this data.
Most notably, Pearson's r assumes that both variables (\ac{LEEP}/\ac{LogME} and post-transfer top-1 accuracy, herein) are normally distributed and have a linear relationship.
Alternatively, weighted $\tau$, a weighted version of the Kendall rank correlation coefficient (Kendall $\tau$), is used in the original \ac{LogME} work.
Weighted $\tau$ is a measure of correspondence between pairwise rankings, where higher performing/scoring models receive higher weight, and only assumes the variables (\ac{LEEP}/\ac{LogME} and post-transfer top-1 accuracy, herein) are continuous.
Both Pearson's $r$ and weighted $\tau$ have a range of $[-1,1]$.
These correlation coefficients confirm the results discussed above.

% Finally, Fig. \ref{fig:leepVlogme} confirms the \ac{LEEP} and \ac{LogME} scores are highly linearly correlated with each other.
% \begin{figure}[t]
%     \centering
%     \includegraphics[width=0.95\linewidth]{figures/leepVlogme.png}
%     \caption{The LEEP versus LogME scores for the sweep over \ac{SNR}, \ac{FO}, and both \ac{SNR} and \ac{FO}. The dashed lines present the linear fit.}
%     \label{fig:leepVlogme}
% \end{figure}

From these figures and metrics it can be concluded that both \ac{LEEP} and \ac{LogME} are strong measures for selecting models for \ac{RF} domain adaptation.
However, as alluded to in the previous subsection, head re-training is more consistent with \ac{LEEP} and \ac{LogME} scores than fine-tuning, as evidenced by higher correlation coefficients. 
Therefore, when using \ac{LEEP} or \ac{LogME} for model selection, using head re-training as a \ac{TL} method would be more reliable than using fine-tuning.
In contrast, fine-tuning, while less reliable than head re-training when used in conjunction with \ac{LEEP} or \ac{LogME} for model selection, offers potential for small performance gains over head re-training.
In practice, this indicates that unless top performance is of more value than reliability, head re-training should be used for \ac{TL} when using \ac{LEEP} or \ac{LogME} for model selection.
In the setting where model accuracy is of the utmost importance, it may be advantageous to try both head re-training and fine-tuning.

It should also be noted that the results shown in Figs. \ref{fig:snr_metricVacc}-\ref{fig:snr_fo_metricVacc} are consistent with the results presented in the original \ac{LEEP} and \ac{LogME} publications where the metrics were tested in \ac{CV} and \ac{NLP} settings, supporting the claim that these metrics are truly modality agnostic.
Therefore, other modality agnostic metrics seem likely to perform well in \ac{RFML} settings as well, and may be examined as follow on work.

%%%%%%%%%%%%%%%%%%%%%%%%%%%%%%%%%%%%%%%%%%%%%%%%%%%%%%%%%%%%%%%%%
\subsection{How can  \ac{LEEP}/\ac{LogME} be used to predict post-transfer accuracy?}\label{sec:pred_method}

Having confirmed that \ac{LEEP} and \ac{LogME} can be used to select models for \ac{RF} domain adaptation, what follows is an approach to not only select models for \ac{RF} domain adaptation, but also to predict the post-transfer top-1 accuracy without any further training. 
The approach is time and resource intensive to initialize, but once initialized, is fast and relatively inexpensive to compute and shows the predictive capabilities of these metrics.

Given $n$ known domains and assuming a single model architecture, to initialize the approach:
\begin{easylist}[enumerate]
@ Run baseline simulations for all $n$ known domains including pre-training source models on all domains, and using head re-training and/or fine-tuning to transfer each source model to the remaining known domains
@ Compute \ac{LEEP}/\ac{LogME} scores using all pre-trained source models and the remaining known domains.
@ Compute post-transfer top-1 accuracy for all transfer-learned models, constructing datapoints like those displayed in Figs. \ref{fig:snr_metricVacc}-\ref{fig:snr_fo_metricVacc}.
@ Fit a function of the desired form (i.e. linear, logarithmic, etc.) to the \ac{LEEP}/\ac{LogME} scores and post-transfer top-1 accuracies. For example, a linear fit of the form $y = \beta_0 x + \beta_1$ is shown in Figs. \ref{fig:snr_metricVacc}-\ref{fig:snr_fo_metricVacc} such that $x$ is the transferability score and $y$ is the post-transfer top-1 accuracy.
@ Compute the margin of error by first calculating the mean difference between the true post-transfer top-1 accuracy and the predicted post-transfer top-1 accuracy (using the linear fit), and then multiplying this mean by the appropriate z-score(s) for the desired confidence interval(s) \cite{hazra2017}.
\end{easylist}

Then, during deployment, given a new labelled target dataset:
\begin{easylist}[enumerate]
@ Compute \ac{LEEP}/\ac{LogME} scores for all pre-trained source models and new target dataset.
@ Select the pre-trained source model yielding the highest \ac{LEEP}/\ac{LogME} score for \ac{TL}.
@ Use the fitted linear function to estimate post-transfer accuracy, given the highest \ac{LEEP}/\ac{LogME} score, and add/subtract the margin of error to construct the confidence interval.
\end{easylist}
Optionally, after transferring to the new labelled target dataset, add this dataset to the list of known domains, and update the linear fit and margin of error, as needed.

\begin{figure}[t]
    \centering
    \includegraphics[width=0.5\linewidth]{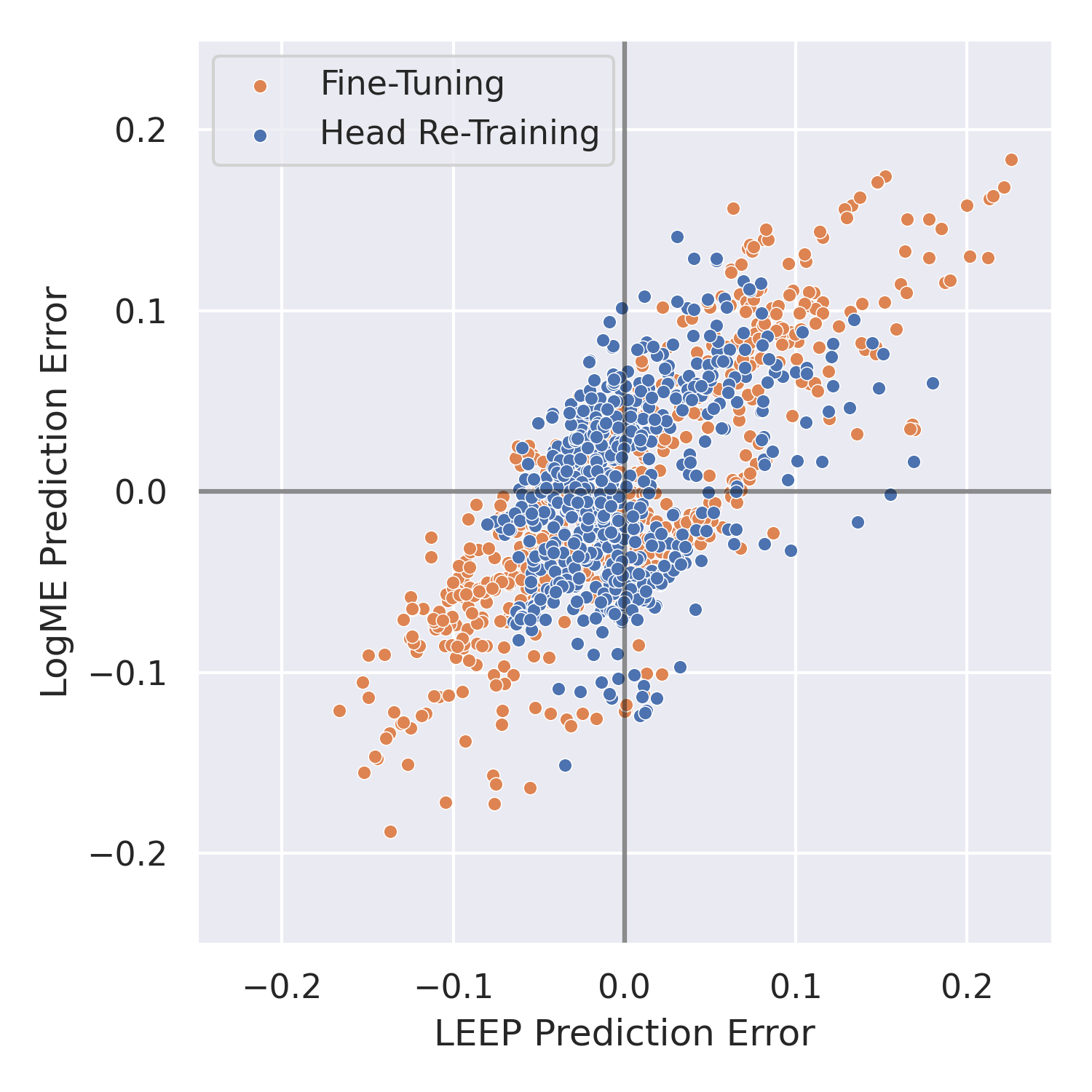}
    \caption{The error in the predicted post-transfer accuracy using a linear fit to the \ac{LEEP} scores (x-axis) and \ac{LogME} scores (y-axis) for the sweep over \ac{SNR}.}
    \label{fig:snr_fiterror}
\end{figure}

\begin{figure}[t]
    \centering
    \includegraphics[width=0.5\linewidth]{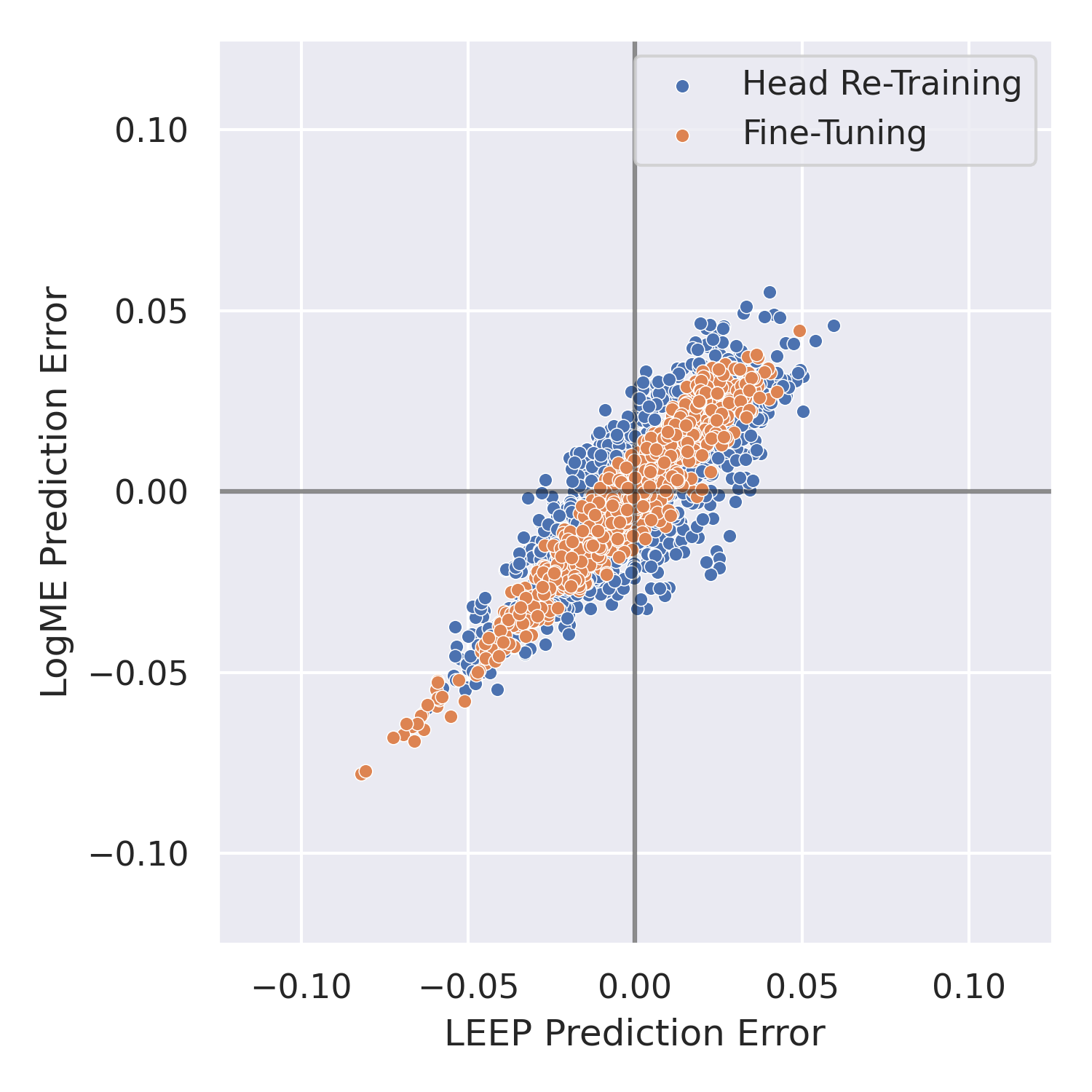}
    \caption{The error in the predicted post-transfer accuracy using a linear fit to the \ac{LEEP} scores (x-axis) and \ac{LogME} scores (y-axis) for the sweep over \ac{FO}. Note the change in scale compared to Figs. \ref{fig:snr_fiterror} and \ref{fig:snr_fo_fiterror}.}
    \label{fig:fo_fiterror}
\end{figure}

\begin{figure}[t]
    \centering
    \includegraphics[width=0.5\linewidth]{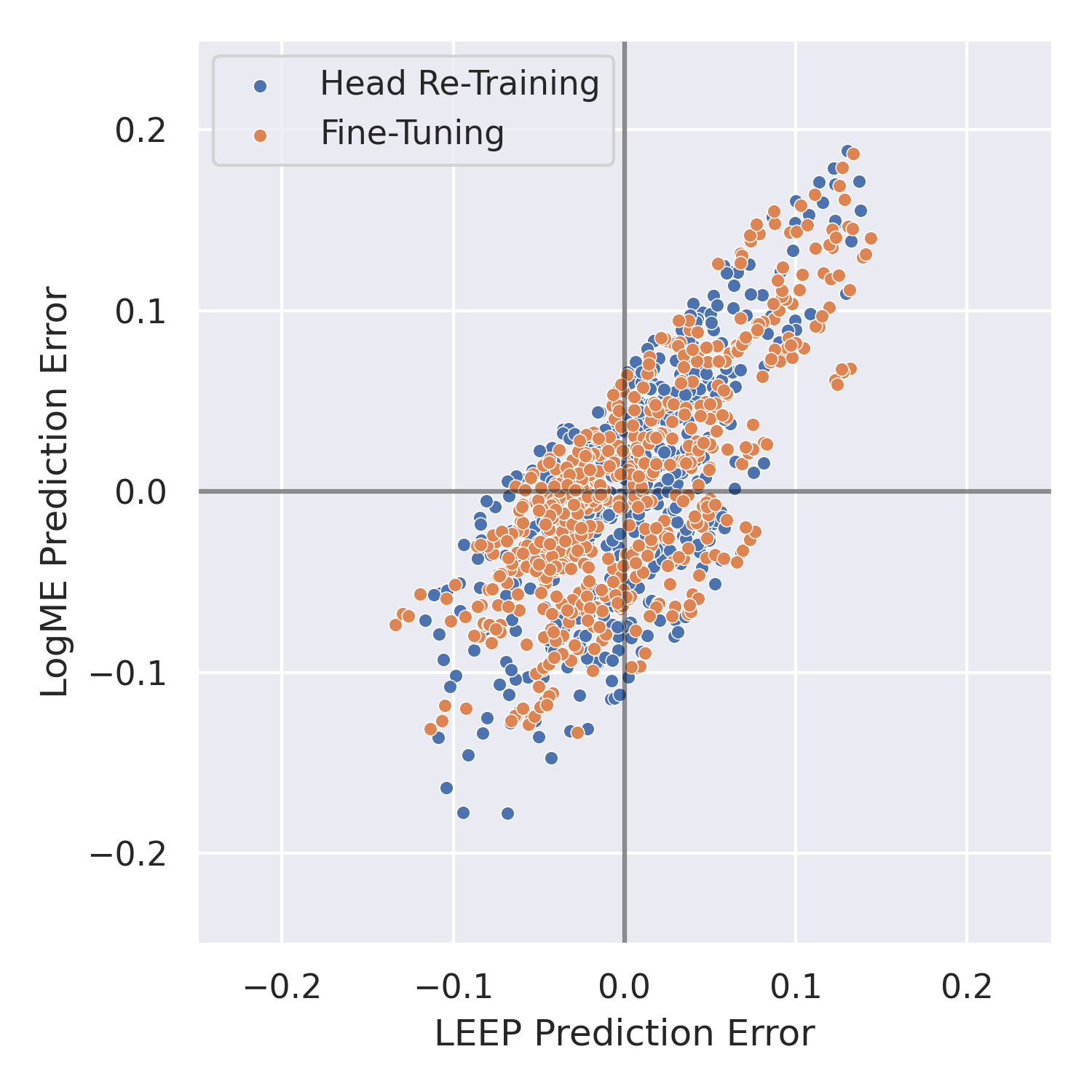}
    \caption{The error in the predicted post-transfer accuracy using a linear fit to the \ac{LEEP} scores (x-axis) and \ac{LogME} scores (y-axis) for the sweep over both \ac{SNR} and \ac{FO}.}
    \label{fig:snr_fo_fiterror}
\end{figure}

The error in the predicted post-transfer accuracy using the proposed method is shown in Figs. \ref{fig:snr_fiterror}-\ref{fig:snr_fo_fiterror}.
These plots show that not only are \ac{LEEP}/\ac{LogME} highly correlated with post-transfer top-1 accuracy (as shown in Figs. \ref{fig:snr_metricVacc}-\ref{fig:snr_fo_metricVacc}), but the error in the predicted post-transfer top-1 accuracy using a linear fit to the \ac{LEEP} and \ac{LogME} scores respectively is also highly correlated.
More specifically, when the proposed method constructed using \ac{LEEP} predicts a lower/higher post-transfer accuracy than ground truth, the proposed method constructed using \ac{LogME} will do the same with the frequencies shown in Tab. \ref{tab:results}.
This indicates that these scores could be combined to create a more robust transferability metric and more robust post-transfer accuracy prediction with relative ease, which is left for future work.

\begin{table}[]
\centering
\caption{The frequency with which the proposed method constructed using \ac{LEEP} and \ac{LogME} agree in over/under predicting post-transfer accuracy.}
\label{tab:results}
\small{
\begin{tabular}{@{}l|ccc@{}}
 & \textbf{SNR Sweep} & \textbf{FO Sweep} & \textbf{SNR + FO Sweep} \\ \midrule
\textbf{Head Re-Training} & 0.6175 & 0.7856 & 0.7258 \\
\textbf{Fine-Tuning} & 0.7496 & 0.8803 & 0.7468
\end{tabular}}
\end{table}

\section{Future Work}\label{sec:fw}
As previously mentioned, several new transferability metrics were developed concurrently with this work, and are suitable as replacements for \ac{LEEP} and \ac{LogME} in any of the above experiments.
Therefore, the first direction for future work is replicating this work using alternative metrics such as \ac{OTCE} \cite{tan2021a}, \ac{JC-NCE} \cite{tan2021b}, TransRate \cite{huang2021}, and \ac{GBC} \cite{pandy2021}, to identify if these metrics are also suitable for use in the context of \ac{RFML} and if these metrics might outperform those used herein.
Given that this work supports the claim that \ac{LEEP} and \ac{LogME} are modality agnostic, it seems likely that additional transferability metrics that are also modality agnostic by design will also follow this trend.
Additionally, the concept of transferability metrics and the experiments performed herein should be extended to inductive \ac{TL} settings including multi-task learning and sequential learning settings in which the source and target tasks differ (i.e. adding/removing output classes), as this work only considered \ac{RF} domain adaptation.

Another direction for future work is the development of new transferability metrics that are more robust than \ac{LEEP} or \ac{LogME} alone or are \ac{RFML}-specific.
Most apparently, results discussed previously in Section \ref{sec:pred_method} indicate that \ac{LEEP} and \ac{LogME} could be combined to create a more robust transferability metric and more robust post-transfer accuracy prediction with relative ease.
However, while modality agnostic metrics such as \ac{LEEP} and \ac{LogME} are shown herein to be suitable for use in \ac{RFML}, a transferability metric purpose-built for the \ac{RF} space would likely be more widely accepted amongst traditional \ac{RF} engineers \cite{wong2021ecosystem}.

With or without the use of transferability metrics, this work provides generalized conclusions about how best to use \ac{TL} in the context of \ac{RFML}.
Provided future verification and refinement of these results and guidelines using captured and augmented data \cite{clark2020}, this work can be used in future \ac{RFML} systems to construct the highest performing models for a given target domain when data is limited.
More specifically, these guidelines begin a discussion regarding how best to continually update \ac{RFML} models once deployed, in an online or incremental fashion, to overcome the highly fluid nature of modern communication systems \cite{wong2021ecosystem}.

\section{Conclusion}\label{sec:conclusion}
\Ac{TL} has yielded tremendous performance benefits in \ac{CV} and \ac{NLP}, and as a result, \ac{TL} is all but commonplace in these fields.
However, the benefits of \ac{TL} have yet to be fully demonstrated and integrated in \ac{RFML}.
To begin to address this deficit, this work systematically evaluated \ac{RF} domain adaptation performance as a function of several parameters-of-interest, including \ac{SNR} and \ac{FO}, using post-transfer top-1 accuracy and existing transferability metrics, \ac{LEEP} and \ac{LogME}.
This work demonstrated that \ac{LEEP} and \ac{LogME} correlate well with post-transfer accuracy, and can therefore be used for model selection successfully in the context of \ac{RF} domain adaptation.
Further, an approach was presented for predicting post-transfer accuracy using these metrics, within a confidence interval, and without further training.

Through this exhaustive study, a number of guidelines have been identified for when and how to use \ac{TL} for \ac{RF} domain adaptation successfully.
More specifically, results indicate:
\begin{easylist}[itemize]
@ Using source models trained on the most similar domain to the target yields highest performance
@ Transferring from a more challenging domain than the target, is preferred to transferring from an easier domain
@ Selecting source models based on this similarity of the source/target channel environment is more important than the similarity of the source/target platform(s)
@ Head re-training is more reliable, faster, and less computationally expensive than fine-tuning for \ac{RF} domain adaptation
@ \ac{TL} via both head re-training and fine-tuning should be attempted, when top performance is of greater value than time and/or computational efficiency
\end{easylist}
These takeaways can be used in future \ac{RFML} systems to construct higher performing models when limited data is available, and can be used to develop to online and active \ac{RFML} approaches, a critical need for practical and deployable \ac{RFML}.

%%
%% The next two lines define the bibliography style to be used, and
%% the bibliography file.
\bibliographystyle{ACM-Reference-Format}
\bibliography{_bibliography}

%%
%% If your work has an appendix, this is the place to put it.
\appendix
\printnomenclature

\end{document}